\documentclass{aa}  
\usepackage{graphicx}
\usepackage{txfonts}
\usepackage{amsmath}
\usepackage{amssymb}
\usepackage{multicol}
\usepackage{natbib}
\usepackage{comment}
\usepackage[colorlinks=true,allcolors=blue]{hyperref}
\newcommand{\ergs}[0]{\rm erg\,\rm s^{-1}}

\newcommand{\gcm}[0]{\text{g}{\,}\text{cm}^{-3}}
\newcommand{\pbb}{\textcolor{blue}{PBRB17}}
\newcommand{\ptw}{\textcolor{blue}{P20}}

\begin{document} 
   \title{Stars as triggers of interstellar gas entrainment in relativistic jets}

    \titlerunning{Stars as triggers of interstellar gas entrainment in relativistic jets}
   
   \author{B. Longo\inst{1}
          \and
          M. Perucho
          \inst{1, 2}
          \and
          J. M. Mart\'{\i}
          \inst{1, 2}
          \and
          V. Bosch-Ramon
          \inst{3}
          \and
          Y. Hoche
          \inst{1}
          }

    \authorrunning{Longo, Perucho, Mart\'{\i}, Bosch-Ramon \& Hoche}
          
   \institute{ Departament d'Astronomia i Astrof\'{\i}sica, Universitat de Val\`encia, Av. Vicent Andrés Estellés
    19, 46100, Burjassot, Val\`encia, Spain.
         \and
             Observatori Astron\`omic, Universitat de Val\`encia, C/ Catedr\`atic Jos\'e Beltr\'an 2, 46980, Paterna, Val\`encia, Spain.
             \and
            Departament de F\`isica Quàntica i Astrof\'isica, Institut de Ciències del Cosmos (ICC), Universitat de Barcelona (IEEC-UB), Mart\'i i Franquès 1, E08028 Barcelona, Spain.
             }
   \date{Received 7 July 2025; accepted 1 September 2026}
  \abstract 
   {Low-power extragalactic jets are known to be decelerated and dissipate large amounts of energy within their host galaxies. However, the exact process by which this occurs is still elusive.}
   {The aim of this work is to probe the role of stars as triggers of jet mass-loading, deceleration and dissipation in Fanaroff-Riley type I radio galaxies. This is motivated by a theoretical model that proposes that stars interacting with the jet boundaries could facilitate entrainment of interstellar medium (ISM) gas into jets, favouring mixing and dissipation.}
   {We have performed a numerical experiment of stars entering a relativistic flow, using a relativistic hydrodynamics code. Our setup is limited to the interaction of three stars with the jet boundary, in order to assess the results in a limited, controlled, environment, although this number of stars may be plausible in the inner kpc-region of a massive galaxy.}
   {Our results allow us to estimate the amount of entrained ISM gas as the stars enter the jet. 
   We show that the entrainment temporally induced on scales of tens of parsecs and thousands of years by evolved stars is comparable to the initial jet mass rate.  
   
   The way in which this entrainment happens is by the creation of a low pressure region behind the stellar objects, which drags ambient gas into the jet flow.}
   {Our results confirm that stars interacting with the jet boundaries, and acting as catalysts of ISM/shear gas entrainment, can significantly contribute to jet mass-load and deceleration.}

   \maketitle

\section{Introduction}

Relativistic jets hosted by active galaxies originate from supermassive black holes (SMBH) in active galactic nuclei (AGN), and are powered by the extraction of rotational energy from the black hole \citep{blandford77,2019ApJ...875L...1E}. From the jet-formation compact regions at sub-parsec scales, the jets can propagate up to distances of hundreds of kiloparsecs or even megaparsecs \citep[e.g.,][]{2020A&A...635A...5D,2024Natur.633..537O,2025A&A...699A.257A}. At parsec scales, jets show collimated structures and relativistic speeds \citep[e.g.,][]{2001ApJ...552..508G,2019ApJ...874...43L}. At kiloparsec scales, however, there is a morphological dichotomy between edge-brightened sources, Fanaroff-Riley type II radio galaxies \citep[FRII,][]{fr74}, and 'edge-darkened' sources, Fanaroff-Riley type I (FRI). Moreover, the former show asymmetrical brightness in jets up to the interaction region with the intergalactic medium (IGM) on hundreds of kiloparsecs and beyond, whereas the latter become symmetric already at kiloparsec scales. Fanaroff-Riley II radio galaxies are, in general, high-power jets and faint radio emitters until they interact with the IGM, where they show a hotspot and lobes in radio \citep[e.g.,][]{1993AJ....105.1690F,1996A&ARv...7....1C,1997AJ....114.2292F,2002ApJ...581..948H}. In contrast, FRI types are low-power and show bright jets at the inner kiloparsec, until they fade into the IGM on scales of tens to hundreds of kiloparsecs \citep[e.g.,][]{2002MNRAS.336..328L,2002MNRAS.336.1161L,2008MNRAS.386..657L,2011MNRAS.417.2789L,laing14}. These differences are interpreted in terms of FRII jets remaining mildly relativistic up to large distances, while FRI jets are decelerated within the host galaxy, thus reducing Doppler boosting.

The current paradigm explains this difference in behaviour by means of entrainment of interstellar medium (ISM) gas \citep{bicknell84,bicknell86}.
However, the driving mechanism for entrainment is still elusive, and, although jet deceleration in FRI radio galaxies has been thoroughly studied, there is still not a clear answer to this point. Among the different possible mechanisms one can find \citep[see][for a review]{perucho19}: entrainment by stellar winds \citep[e.g.,][]{komissarov94,bowman96,hub06,perucho14,angcast21,fichet25}, strong recollimation shocks that could decelerate the flow and make it prone to the development of Kelvin-Helmholtz instabilities \citep[e.g.,][]{perucho07}, sharp variations in the properties of the ambient medium \citep[e.g.,][]{2008A&A...491..321M}, the development of large-scale \citep[e.g.][]{2008A&A...488..795R,2016A&A...596A..12M,2019A&A...621A.132M,2020A&A...633L...1R,2022A&A...659A.139M,2024A&A...685A...4R} or small-scale instability modes driven by rotation \citep[e.g.][]{2007A&A...475..785M,2009ApJ...705.1594M}, velocity shear \citep[e.g.][]{perucho10}, expansion \citep[e.g.,][]{matsumoto17,gourg18a}, or stars crossing the jet-ambient separation surface \citep[][\ptw~from now on]{perucho20}. In \citet{laing14}, the authors established an observational paradigm, showing evidence of the co-spatiality of deceleration and brightness enhancements, which led them to conclude that the process takes place through strong dissipation, accompanied by significant radiation losses. Furthermore, that work also showed that deceleration is progressively affecting the whole jet cross-section, from the jet boundary to its axis, while the jets do not show any large-scale deformation other than expansion.

Taking this observational evidence into account, only the scenarios involving small-scale processes seem to pass the test. According to \citet{matsumoto17,gourg18a}, this could be produced by the development of Rayleigh-Taylor/centrifugal instability modes generated by jet expansion and recollimation. However, the canonical FRI jets observed by \citet{laing14} do not show hints of recollimation or strong shocks. Relying on the fact that the jets are expected to contain millions of stars \citep[e.g.,][]{wykes15}, \ptw\, proposed stars crossing the jet boundaries as a way to trigger entrainment by the ISM, mainly via the excitation of instabilities. This model was inspired by numerical simulations of a star crossing the jet boundary, which showed that it could perturb the jet surface, and also drag ISM gas into the jet \citep[][\pbb~from now on]{perucho17}. According to \ptw\, one such interaction every few hundred years per parsec along the innermost one hundred parsecs from the nucleus could be enough to develop a mixing layer. Such a layer could then propagate towards the jet axis at about the local sound speed. 

Therefore, stars could play a dual role in jet dissipation and deceleration. On the one hand, medium gas can be efficiently entrained as stars enter the jet if they act as mediators of jet-ISM mixing or trigger the development of instabilities (\ptw). On the other hand, it can occur as well via the direct entrainment of protons and nuclei injected by the stellar wind in the flow \citep[e.g.][]{komissarov94,bowman96,perucho14,angcast21,fichet25}. It should be also noted that since strong dissipation of jet energy can occur at both the bow shock generated at the jet-wind interaction and the downstream mixing region, efficient particle acceleration and non-thermal radiation can take place \citep[e.g.,][\pbb]{bosch12,araudo13}. 

In this work we perform an initial test of the model proposed by \ptw~and evaluate the role of stars as triggers of ambient/ISM entrainment. With this aim, we present a numerical experiment limited to the interaction of a few stars with a relativistic jet. The stars are represented as clouds of gas with the size and density of the shocked wind region (see \pbb). We study the entrainment of ISM gas driven directly by the stellar crossing, and the effect of the stars on the jet boundary as a plausible origin of instabilities developing farther downstream. This limits our simulation to an intermediate scale, in which we do not focus on the details of jet-star interaction or on the development of instabilities downstream of the interaction region. On the one hand, spatial scales in which the jet-wind interaction takes place are typically much smaller than the jet radius. Therefore, we do not aim to resolve the jet-wind interaction in detail, as our stellar objects themselves involve only a few cells. This implies that the mass-load by the stellar wind is not properly resolved in the simulation, and thus we separate the stellar component from the entrained ISM component in our analysis. On the other hand, the scales for the development of instabilities and a mixing layer are much larger than the jet radius, implying that a full study of this development would require very large grids and, consequently, massive computational resources. 
Nevertheless, at the present approximation level, our results seem to confirm the possibility that stars with significant stellar wind bubbles/interaction regions can strongly contribute to direct mass-loading of the jet with entrained ISM and shear gas and certainly perturb the jet shear layer. 

The paper is structured as follows: In Section~\ref{setup}, we describe our setup both from an astrophysical and a numerical perspective; in Section~\ref{results} we present our results and analysis derived from our simulations; in Section~\ref{disc} we discuss those results; and in Section~\ref{sum} we summarise our results and give our conclusions. 

\section{Setup} \label{setup}

\subsection{Simulation}\label{simulations_ref}

To run the simulations, we used the hybrid MPI+OpenMP finite-volume code \texttt{RATPENAT} \citep{perucho10}, which solves the relativistic hydrodynamics (RHD) equations in conservation form by means of high-resolution shock-capturing methods (HRSC). The code uses the Synge equation of state \citep{synge57}, which allows us to describe a gas composed of (relativistic) protons, electrons and positrons. The conservation equations also include the evolution of a passive scalar,
$f$, which gives the jet-mass fraction and allows us to trace the mixing between the jet ($f=1$) and the ISM-stellar medium ($f=0$).

\begin{figure}
\centering
\includegraphics[width=\linewidth]{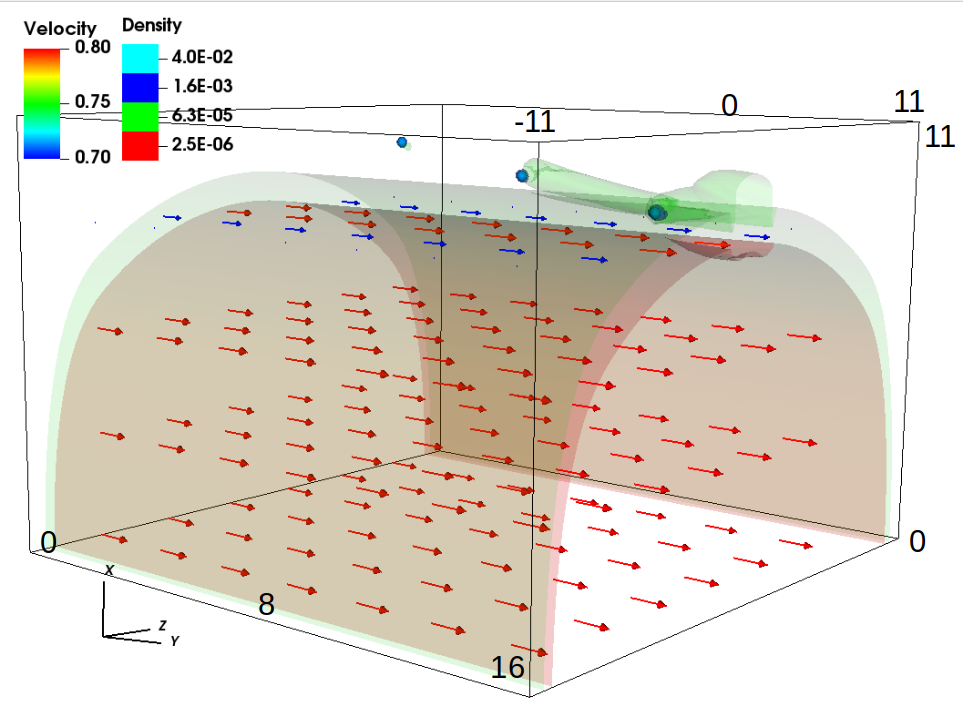}
\caption{The figure shows the simulation setup (see the text) close to $t=0$, when two of the stars, S1 and S2, have initiated the interaction with the jet shear-layer. The contours show constant rest-mass density surfaces, whereas the axial velocity is indicated by coloured arrows.}
\label{initjbs3}
\end{figure}

We model the jet as half a cylinder (see Fig.~\ref{initjbs3}), 
cut in the $yz$-plane, with the jet flow in the $y$ direction. The grid covers the physical domain $[0,110]\,R_{\rm b}\times[0,160]\,R_{\rm b} \times [-110,110]\,R_{\rm b}$ (where $R_{\rm b}=0.1\,{\rm pc}$ is the stellar bubble radius) in the $x$, $y$ and $z$ directions respectively ($[0,11]\,{\rm pc}\times[0,16]\,{\rm pc}\times[-11,11]\,{\rm pc}$ in physical units) and a total of $660 \times 960 \times 960$ cells. 

In the $z$ direction, the grid consists of 660 uniform cells covering the central interval $[-55,55]\,R_{\rm b}$ embedded by an extended mesh of 150 + 150 cells whose resolution is progressively degraded by means of a stretching factor $f_{\rm s}=1.009$ from the base cell size, covering the intervals $[-110,-55]\,R_{\rm b}$ and $[55,110]\,R_{\rm b}$. The grid cell size is uniform along the $x$ and $y$ directions.

The resulting resolution in the uniform grid is six cells/$R_{\rm b}$. The jet enters the grid from the inflow boundary at $y=0$, whereas every other boundary is set to outflow conditions. Figure~\ref{initjbs3} shows a snapshot of the simulation close to the initial setup,
with the first two stars S1, of coordinates (9,13.33,0)\,pc, and S2, of coordinates (9.66,9.38,0)\,pc, already interacting with the jet shear layer and developing a cometary tail. The third star, S3, of coordinates (10.33,5.43,0)\,pc, is the closest to the inflow boundary and the last to enter the jet.

The stars are initially located in such a way that the impact they will have on the shear-layer boundary is continuous as their entrance is sequential. Although the upstream objects produce bow-shocks that shield the downstream objects against the jet flow, we use this approach as an illustrative case for a first experiment under controlled conditions, which is also constrained by limited computational resources and resolution. They propagate towards the jet interior at a speed of $\sim 6.6\times10^{-4}\,c$ (see next section). The simulation followed the propagation of the stars during $\sim 10^{4}\,{\rm yr}$, when they have travelled $\sim 2\,{\rm pc}$ (i.e., 1/5 of the jet radius) towards the jet axis in our configuration. This was determined by the time the bow-shocks reach the plane $x=0$.

With a time-step of $\sim 1.5\times10^{-2}\,{\rm yr}$, we have to run the code through $\sim  8\times10^5$ iterations, which is computationally expensive with the mesh used. Therefore, we applied for computing time and were granted $8000\,{\rm kh}$ in the MareNostrum5 supercomputing facility at the Barcelona Supercomputing Center, within the Spanish Supercomputing Network (RES). The simulation was run using 3584 cores.

%
%

\subsection{Stars, ISM and jet properties}
\label{sijp}

Our aim is to study the role of stars as a trigger of jet-ambient medium mixing and not so much as a source of mass-loading themselves. This is in contrast to previous works in which we focused on the jet mass-loading by stellar winds. For this reason, we simulate stars as a moving boundary condition through the grid. The stars move into the jet during their orbital motion around the galactic centre. We set an orbital velocity of $v_{\rm orb}=200\,{\rm km}\,{\rm s}^{-1}$ \citep{mistele24}, and place them outside the jet. We have added a smooth transition (shear layer) of density and velocity between the jet and the ambient medium to keep radial equilibrium and avoid numerical noise or upstream propagating waves (see \pbb). 

Regarding the jet, we model it as a mildly relativistic electron/proton gas with radius $R_{\rm j}=10\,{\rm pc}$ and velocity $v_{\rm j}=0.8\,{c}$. By imposing a density of $4.175\times10^{-28}\gcm$ and a temperature $T_{\rm j}=10^{11}\,$K, we obtain a jet power of $L_{\rm j}=3.43\times10^{43}\,\ergs$,\footnote{with $L_{\rm j}=\rho_{\rm j} W_{\rm j} (h_{\rm j} W_{\rm j} -1) A_{\rm j} v_{\rm j}$, where $W$ is the Lorentz factor, $h$ is the specific enthalpy and $A$ is the jet surface.} a value within the expected FRI power values. The shear layer naturally establishes a central region where the jet parameters are constant, with radius $\sim 8.4\,{\rm pc}$ (which we call the jet \emph{spine} in our analysis) and the transition region to the ambient values.

%
The interaction of stars with their surroundings (ambient medium and shear layer/jet system) leads to the inflation of bubbles of shocked stellar-wind gas (see \pbb; \citealt{torres-alba19}). As the star crosses the jet, the interaction is expected to evolve towards a quasi-steady configuration, with the equilibrium point located at a distance $R_{\rm eq}$ from the star in the direction of the jet flow,
\begin{equation}
    R_{\rm eq} = \sqrt{\dfrac{\dot{M}_w v_w}{4 \pi \rho_{\rm j} h_{\rm j} W_{\rm j}^2 v_{\rm j}^2}} \,,
\end{equation}
set by the balance between jet and stellar-wind momenta \citep{komissarov94}. In the previous expression, $\dot{M}_w$ is the stellar wind mass-loss rate, $v_w$ is the wind velocity, and the parameters in the denominator are jet rest mass-density, $\rho_{\rm j}$, enthalpy, $h_{\rm j}$, Lorentz factor, $W_{\rm j}$, and velocity, $v_{\rm j}$.

For the jet parameters used in this paper, we obtain the following equilibrium distance:
\begin{equation}
    R_{\rm eq} \sim 10^{-2} \left(\dfrac{\dot{M}_{\rm w}}{10^{-5} M_\odot {\rm \,yr^{-1}}}\right)^{1/2} \left(\frac{v_{\rm w}}{100 \, {\rm km\,s^{-1}}}\right)^{1/2}\,{\rm pc}.
\end{equation}
Nevertheless, given that the actual jet-wind interaction structure is larger than the equilibrium radius \citep[the initial size of a wind bubble is expected to be $R_{\rm b}\sim (v_{\rm orb}/v_{\rm w})\,R_{\rm eq}$, i.e., a few times $R_{\rm eq}$;][]{torres-alba19}, and the most evolved stars may approach mass-loss rates of $\sim\,10^{-4}\,$M$_\odot\,$yr$^{-1}$, our calculations already illustrate the relevance of the simulated processes.

In our simulations, stellar-wind bubbles are set up as homogeneous spheres located in the plane defined by $z=0$ (see Fig.~\ref{initjbs3}), with radius $R_{\rm b}$ ($0.1\,{\rm pc}$) and density $\rho_{\rm b} = 4.175\times10^{-21}\,\mathrm{g\,cm^{-3}}$. The gas temperature is set to $T_{\rm b}=10^{4}\,\mathrm{K}$. This setup is similar to that adopted by \pbb, albeit with some relevant differences. First, the sphere is homogeneous, in contrast to the dense inner core surrounded by a decreasing-density envelope used in the previous work. Second, the gas within the sphere is not assigned a radial (wind) velocity; in \pbb, the core of the stellar region was treated as an inner boundary condition with an imposed radial outflow. And more importantly: star wind bubbles are reset after every time-step to their original values to avoid their artificially fast disruption (on a timescale of years, in contrast to the expected $\gtrsim 10^3$~yr; \pbb) due to our present resolution constraints\footnote{This is a known situation in simulations of jets interacting with clumpy media \citep[see, e.g.][]{2021AN....342.1171P,2024A&A...684A..45P} where low resolution unrealistically enhances mixing and heating of shocked bubbles.}. To recover the stellar bubbles in pressure balance with the ISM and the jet as they are in the initial setup, we reset the structure of density, pressure, velocity and tracer $f$ around their current location and, as such, all the remaining quantities.

Keeping the integrity and coherence of the stellar bubbles allows us to study their role in facilitating jet--ambient mixing, at the risk of overestimating the expected stellar entrainment. By artificially recovering the mass lost by the bubble, we increase its mass and thus the jet mass load it produces along the simulation. However, we can quantify and thus control the consequences of this numerical treatment. Finally, we chose to introduce three stellar bubbles within the simulated volume. Driven by the aforementioned limitations of the simulations, and taking into account that the size and masses of the bubbles would correspond to massive, evolved stars, this implies a somewhat large (but still plausible) number density, even for the galactic centre. These aspects and their implications are discussed in Sect.~\ref{disc}.

The ISM is defined as the region that fills the space outside the jet and the stellar-wind bubbles. Here, we impose a density ratio $\rho_{\rm ISM}/\rho_{\rm b}=10^{-4}$ with the stellar-wind bubbles, resulting in $\rho_{\rm ISM} = 0.25 \,m_{\rm p}\,{\rm cm}^{-3} = 4.175\times10^{-25}\,\gcm$. The ISM temperature ($T_{\rm ISM}=10^{8}\,$K) is chosen to force pressure equilibrium between the ISM, the stars and the jet. Pressure balance is imposed to avoid waves or expansion of the stellar bubbles.
Although the ambient temperature is larger than the expected one ($\sim 10^{6-7}\,{\rm K}$), we should note that the medium surrounding the jet in our simulation represents a slow and relatively hot wind, as part of the jet shear-layer generated, e.g., by friction and dissipation.

\section{Results} \label{results}

\subsection{Evolution} \label{evol} 

%
\begin{figure*}[h]
\centering
\includegraphics[width=0.9\linewidth]{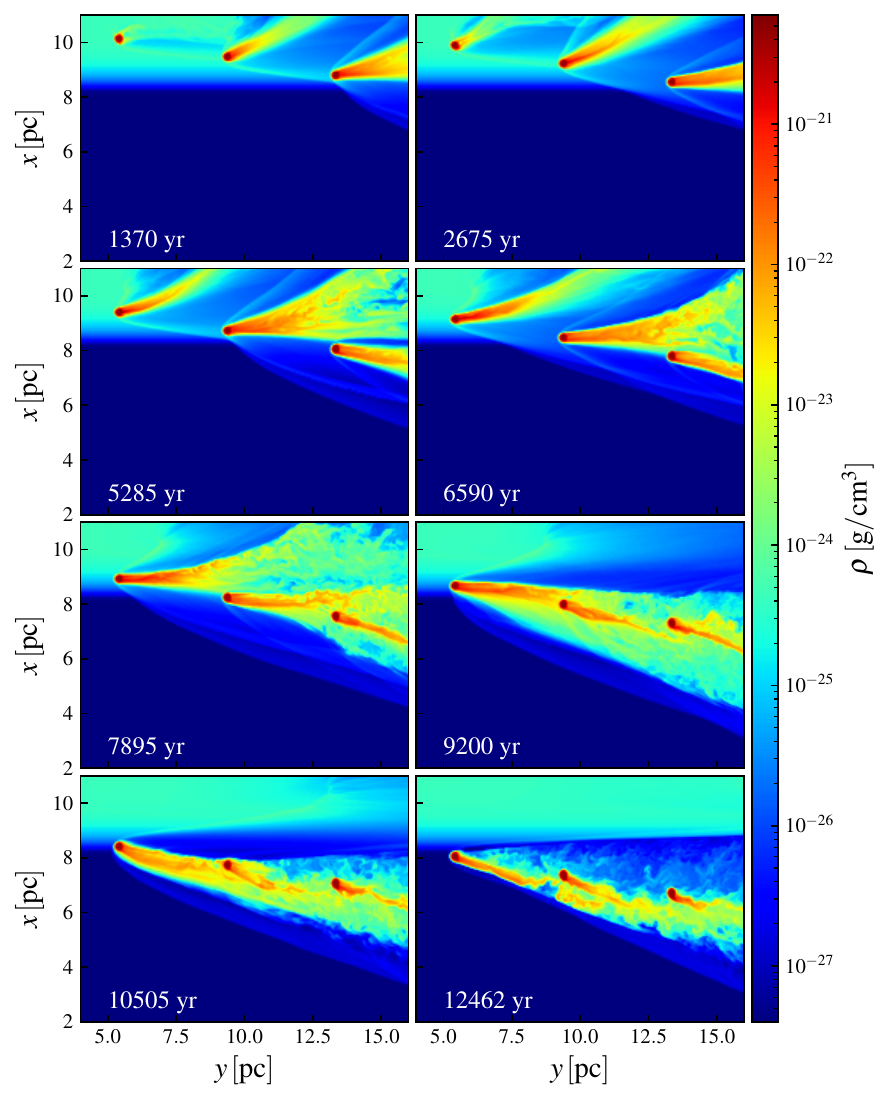}
\caption{Rest-mass density in the $x-y$ plane at $z=0$ at different times along the simulation.}
\label{8snaprho}
\end{figure*}
%
%
\begin{figure*}[t]
\centering
\includegraphics[width=0.75\linewidth]{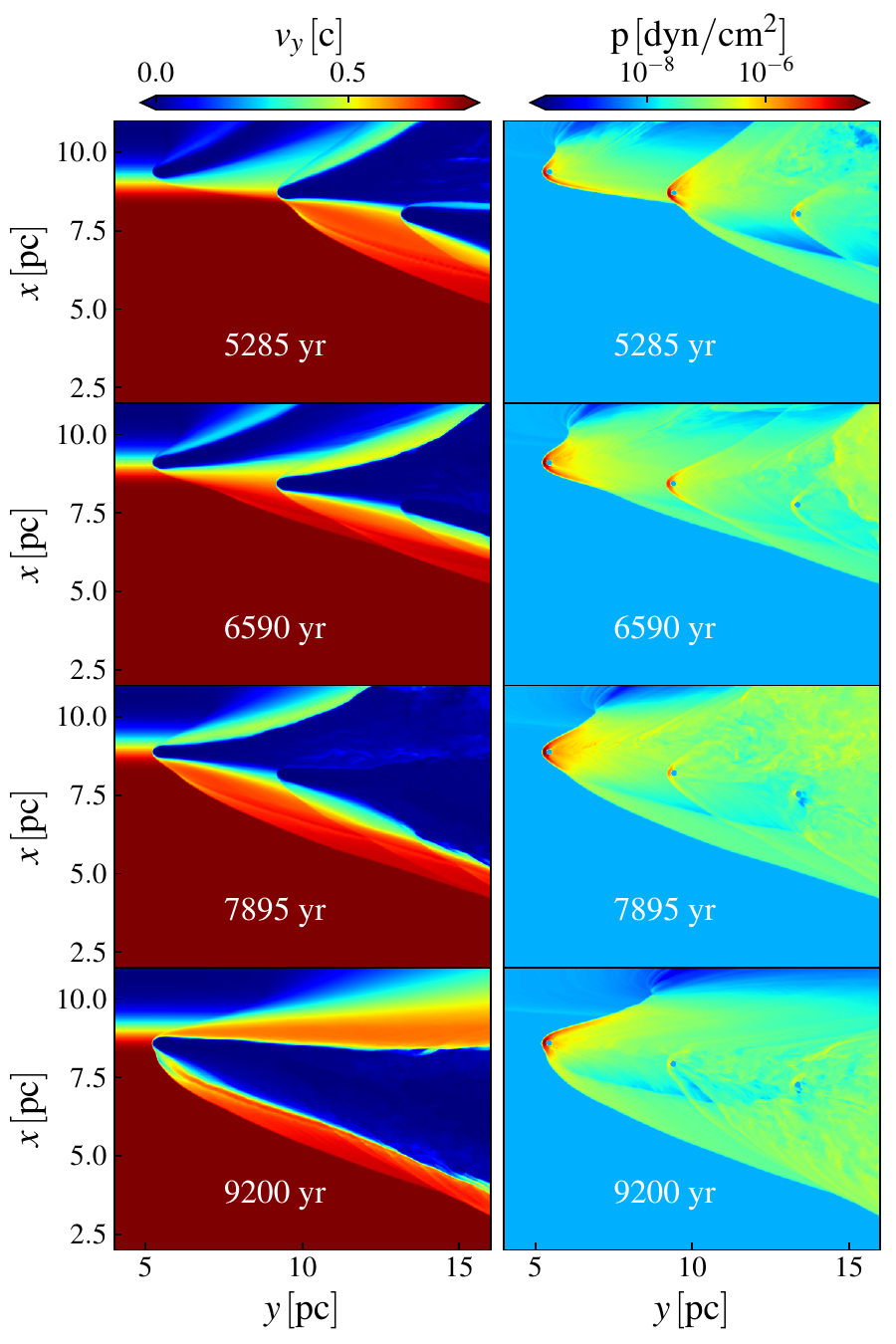}
\caption{Axial velocity (left column) and pressure (right column) in the $x-y$ plane at $z=0$. The selected times coincide with the second and third rows of Fig.~\ref{8snaprho}, i.e., the central part of the simulation.}
\label{8snapprvel}
\end{figure*}
%

Figure~\ref{8snaprho} shows rest-mass density cuts of the jet at different times along the simulation. The stars propagate downward in the image as they enter the jet.
We set three stellar bubbles in $\sim 7000\,$yr, i.e., approximately one every $\sim 2000\,$yr. The first two frames of the Figure (top panels) show fully developed bow shocks for S1 and S2 (rightmost and central stars, respectively), due to interaction with the shear-layer. The third stellar bubble, S3, crosses the shear-layer between $t=5280\,{\rm yr}$ and $t=6590\,{\rm yr}$ (third and fourth panels). 

The plasma ablated from the stellar bubbles forms cometary tails behind the stars (red/yellow/light blue regions behind the bubbles in the panels of Fig.~\ref{8snaprho}). These tails show a component along the $x$ direction as stars propagate across the ISM but change their orientation gradually towards the $y$ direction as they enter the jet. Overall, the bow shocks surrounding the tails undergo the same change of orientation along the star propagation. It is interesting to note the asymmetry in the bow shock angle with the flow direction once the stars are inside the jet. The angle of the bow shock depends on the flow Mach number, with higher Mach numbers leading to narrower angles with the $y$ axis. In our simulations, the shear layer is formed by a colder and denser (as compared with jet) gas and, therefore, the flow has a high Mach number. As a consequence, the upper side of the bow shock appears almost aligned with the jet boundary (see last panel of Fig.~\ref{8snaprho}). However, it is foreseeable that the bow shocks take on a more symmetric shape as the stars approach the jet axis.
It is also remarkable that once inside the jet, the bow shocks of the stars located upstream shield those located downstream from the jet flow: at $t= 2675$ yr (second panel in Fig.~\ref{8snaprho}) the bow shock of S1 is already embedded within the bow shock of S2, and at $t=6590$ yr (fourth panel), the bow shocks of S1 and S2 are embedded in that of S3. This is an interesting feature of the jet/stellar wind interaction process: stars located downstream (S2 and S1 in this case) are not in direct interaction with the jet flow, but with previously shocked jet material or even with a jet/ambient/stellar material turbulent mixing region. The effect of the shielding could be the delay of the complete erosion of the stellar wind bubble, favouring that loading takes place deeper into the jet, and not only at the shear-layer. This effect, which is only relevant for obstacles entering the jet approximately aligned in the $y$ direction, is expected to be more efficient for low-power jets, in which the bow shocks are wider due to the jet lower momentum fluxes.  

Figure~\ref{8snapprvel} shows the axial flow velocity (left panels) and pressure (right panels) distributions for the same cut as in Fig.~\ref{8snaprho}, at $t = 5285, 6590, 7895$ and 9200~yr. In the left panels, stellar and ambient material entrained in the jet fill the deep blue regions, representing barely relativistic axial speeds ($<0.2 \, c$). In the right panels, the increase of pressure behind the bow shocks is clearly seen, with the largest pressures found at the tips of the shocks, specially the leading one.

As mentioned previously, the cometary tails change their orientation gradually from the $x$ to the $y$ direction as they enter into the jet. However, the asymmetry in the leading bow shock changes the direction of the postshock flow beneath the tails towards the jet axis causing a drop in pressure in this region (light blue regions in the pressure panels at $7895$ and $9200$~yr in Fig.~\ref{8snapprvel}). This makes the tails continue falling towards the jet axis and aligning with the lower side of the bow shock in the images (see last panel of Fig.~\ref{8snaprho}).

%

%
\begin{figure*}
\centering
\includegraphics[width=0.8\linewidth]{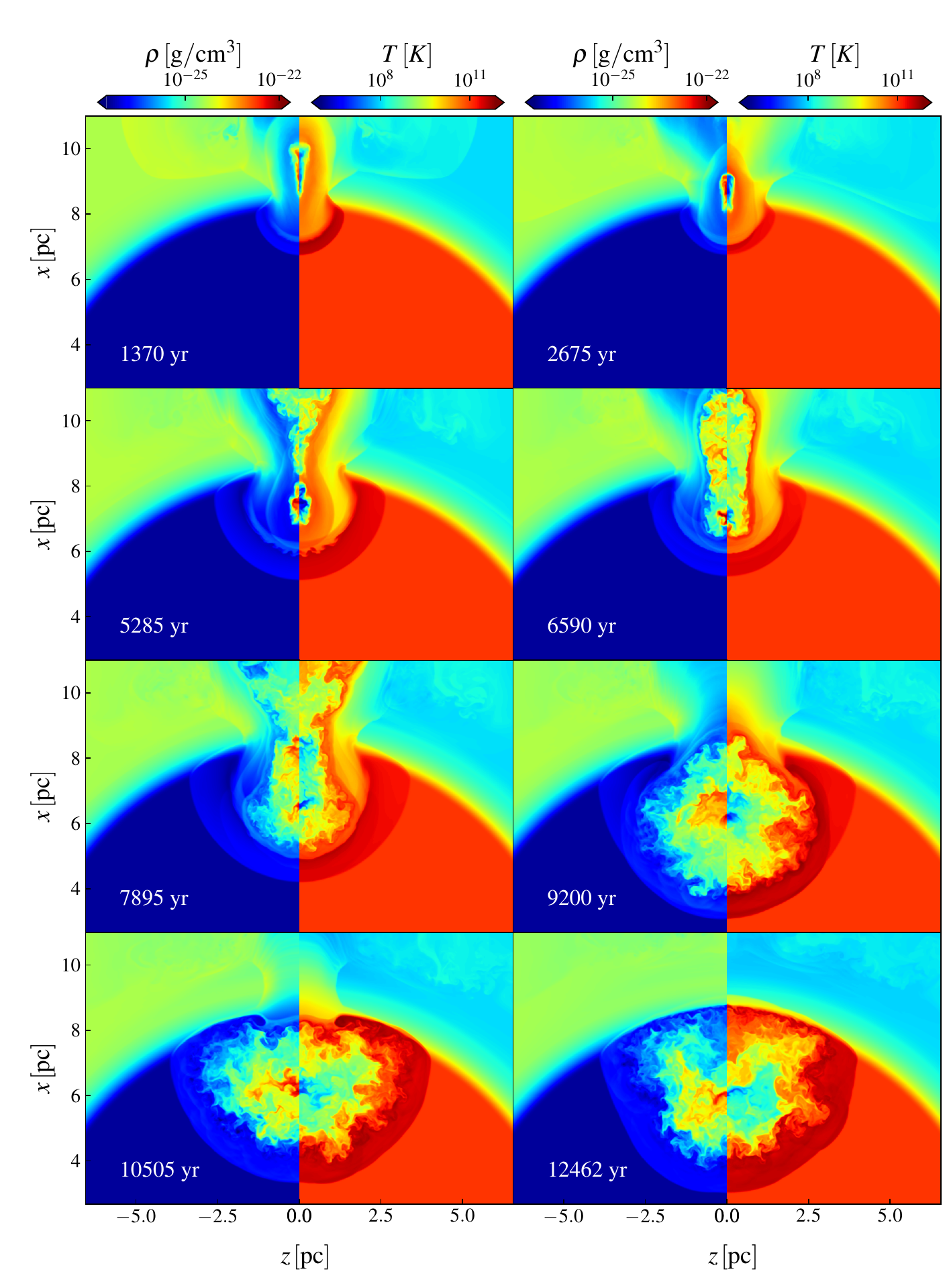}
\caption{Rest-mass density (left half of each panel) and temperature (right half of each panel) in the $x-z$ plane at the downstream boundary ($y=16$pc) at different times along the simulation, coinciding with those shown in Fig.~\ref{8snaprho}.}
\label{8snaprhot}
\end{figure*}
%

Figure~\ref{8snaprhot} displays $x-z$ cuts, transverse to the jet flow, at the downstream boundary ($y=16$~pc), showing the rest-mass density (left half of each panel) and temperature (right half) at the same times as in Fig.~\ref{8snaprho}. The snapshots clearly show the correlation between the denser regions (shown in yellow to red in the density plots) and the colder regions (shown in turquoise to blue in the temperature plots), which are filled with gas from the stellar tails. At the opposite extreme, the more dilute regions (shown in green in the density plots) correspond to hotter gas (green to yellow in the temperature plots) associated with ISM/shear material dragged along by the stars. The ambient gas is entrained by a combination of two main effects: 1) the diversion of shear-layer flow into the jet and 2) a \emph{vacuum-cleaner} (VC) effect operating in the wake of the stellar bubbles: the low-pressure region behind the stars, together with the resulting pressure gradient directed towards the jet interior, pulls this gas into the jet. However, the exact contribution of each of these processes is difficult to assess using our current set up, because there is no discriminating magnitude that can be used in this case. Nevertheless, both processes contribute to entrainment of hotter shear/ISM gas and can be accounted as operating together.

\begin{figure*}
\centering
\includegraphics[width=0.85\linewidth]{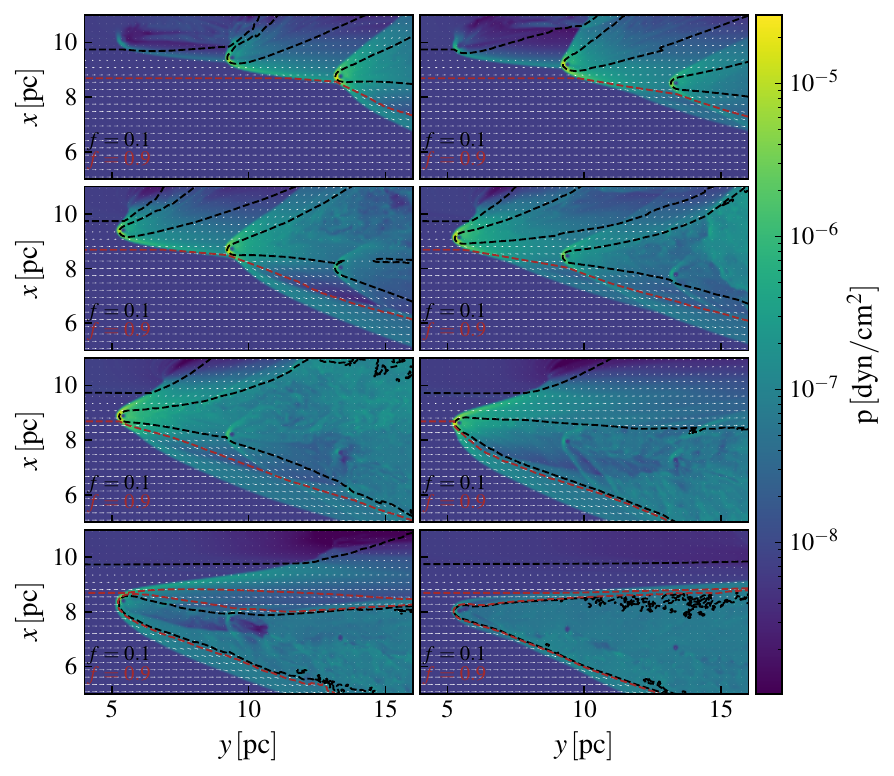}
\caption{Pressure maps in the $x-y$ plane at $z=0$ at the same times along the simulation as shown in 
Fig.~\ref{8snaprho}. The dashed lines indicate the contours for tracer $f=0.1$ (black lines) and $f=0.9$ (dashed lines).}
\label{vc}
\end{figure*}

To illustrate this combination more clearly, Fig.~\ref{vc} shows the logarithm of pressure (colour scale) together with the velocity field. In the panels, two dashed lines indicate the levels of tracer 0.9 (red) and 0.1 (black). We see that during the crossing of the shear layer, this material, denser than the jet flow but with non-zero velocity, is diverted by the shock towards the jet interior. Its temperature is also high, as corresponds to the transition between the jet and the hot ISM, and it can be identified with the red stripe of material following the bow-shock shape into the jet in the left column panels of Fig.~\ref{8snaprhot}. Regarding the VC effect, there are two fundamental ingredients to be taken into account: 1) a general drop in pressure is produced behind the obstacles, and 2) the shocked material is trapped behind the bow shocks. When the stars are completing the crossing of the shear layer, hot, shocked jet material can be deviated towards the shear layer, heating up the region. This process involves shocked ambient gas via mixing. Finally, the gas in this overpressured region propagates into the jet driven by the strong pressure gradient, and accompanied by the shock-obstacle motion. This is indicated by the arrows pointing in the down-right direction and the shape of the interaction region in the bottom-left panel of Fig.~\ref{vc}.

The last snapshots of Figs.~\ref{8snaprho} and \ref{8snaprhot} show that the jet tends to recover its original cylindrical boundary layer. However, its interior has evolved from the laminar flow structure seen in the top-left panels of these figures into a complex and turbulent flow, driven primarily by shocks and by the mixing of jet material with both stellar and ISM/shear gas. Mass entrainment from the hot ISM and the shear-layer through the previously described effects may continue for some time, as long as the jet-star bow shocks still reach the jet/ISM boundary. However, the temporal and spatial extent of our simulations does not allow us to follow this process over its full development.
%
\begin{figure*}
\centering
\includegraphics[width=.8\linewidth]{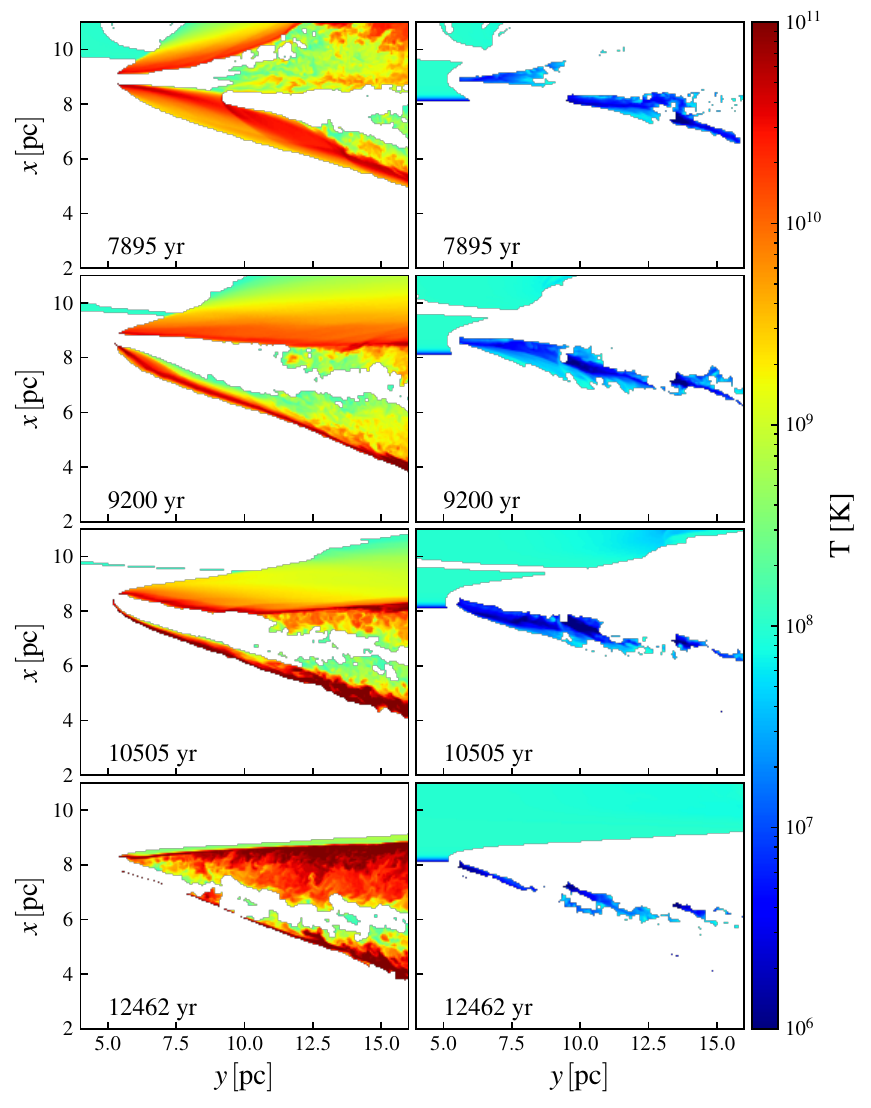}
\caption{Temperature colour maps of the entrained ambient (left column) and stellar wind (right column) gas in the $x-y$ plane at $z=0$. The selected times coincide with third and fourth rows of Fig.~\ref{8snaprho}, i.e., the last part of the simulation, when entrainment is maximum. The ambient gas has been characterized by using the tracer $f<1$ and temperatures $T>10^8$~K, whereas the stellar wind is identified with the tracer $f<1$ and temperatures $T<10^8$~K (see the text).}
\label{8snaprhoft}
\end{figure*}
%

%
%

\subsection{Entrainment and dissipation} \label{ed}

One of the aims of this experiment is to assess the amount of ISM/shear gas entrained by the jet during the stellar penetration process. For this, we need to separate the entrainment of gas from the stellar bubbles, which is affected not only by the artificial refilling but also by the low resolution used to resolve them (6 cells per radius). We can use the gas temperature to distinguish both components. Figure~\ref{8snaprhoft} shows cuts of the density distribution of entrained material (obtained by weighting the density with the tracer, $1-f$)
separated into the hot (ISM/shear) component, with $T>10^8$~K (left panels), and the cold, stellar component, with $T \leq 10^8$~K (right panels). 
This identification is based on the fact that the stellar bubble material is initially at $10^4$~K, whereas the ISM/shear component is at $10^8$~K. Considering that shocks and mixing heat up the gas, we expect this hot component to attain temperatures higher than the initial value (see also Fig.~\ref{8snaprhot}). Although the cold, stellar component may also be heated,
we do not expect a large fraction of it to reach $10^8$~K within the computational grid. Although mixing between the gas extracted from the bubbles and jet or shocked ISM gas can produce a leakage of stellar material to the hot component, the extension of the stellar tails and the expected stability of cold, dense flows allows us to be confident that this threshold gives a robust criterion to distinguish between the two components. Nevertheless, we plan to introduce specific tracers for the ISM and the stellar material in order to alleviate this uncertainty and verify our results in future work. The plots show that the material stripped from the stellar bubbles forms stable tails (blue regions in the right panels), as expected from their relatively large densities \citep[][]{bosch12,perucho17}. These tails are surrounded by hotter (orange/red regions in the left panels), lower-density gas entrained from the ISM/shear as a result of stars piercing and penetrating the jet. 
%
\begin{figure*}
\centering
\includegraphics[width=0.48\linewidth]{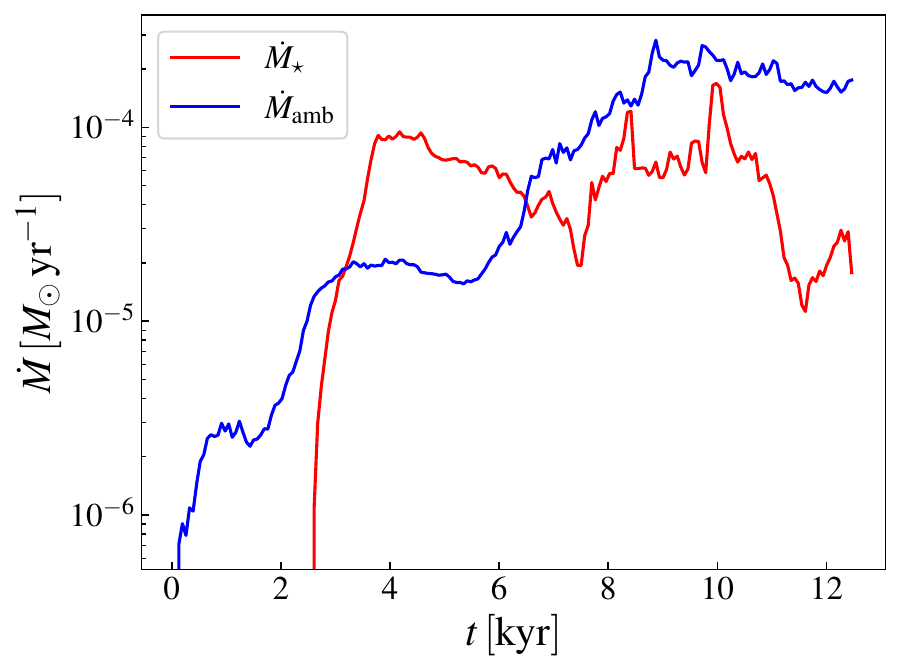}
\includegraphics[width=0.505\linewidth]{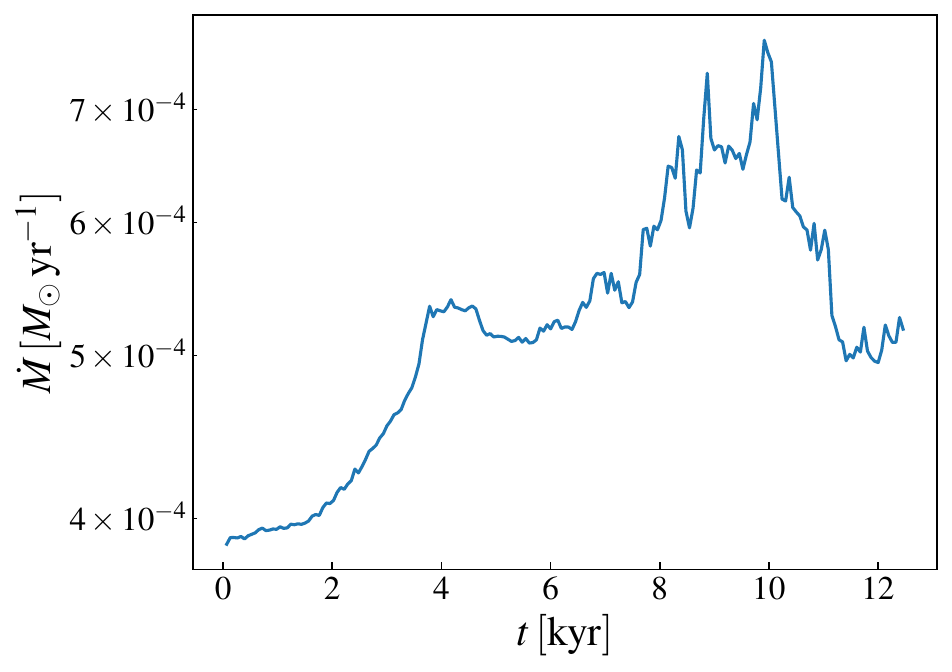}
\caption{Total mass flux across the downstream outflow boundary
within a radius of 8.4~pc (i.e., the spine), as a function of time.
The left panel shows the ambient (blue line) and stellar (red line) components of the mass flux (see the text). The right panel shows the total mass flux, which includes the jet contribution. For reference, the total mass flux carried along the simulated jet spine at injection is $3.5\times 10^{-4}\,M_\odot {\rm \,yr^{-1}}$ (and twice this value if we consider the whole jet). Stars S1, S2 and S3 reach the jet interior at $t \approx 2600, 5300$ and 8000~yr, respectively.}
\label{mload}
\end{figure*}
%

Basing on the same approach, the left panel of Fig.~\ref{mload} shows the contributions of both the colder stellar component (red line) and the hotter ISM/shear component (blue line) to the total total mass flux\footnote{$\dot{M} = A\rho W v_y$, where $A$ is the jet surface, $\rho$ is the rest-mass density, $W$ is the jet Lorentz factor, and $v_y$ is the axial velocity.} through the downstream ($y$-direction) outflow boundary as a function of time. We restrict the calculation to the jet spine, with a radius of 8.4~pc, to avoid the influence of the shear layer, which contributes significantly to the mass flux. Integrating the curves over time yields a total entrained mass of $0.75\,{M_\odot}$ for the stellar component, and $1.25\,{M_\odot}$ for the ISM/shear component throughout the whole of the simulation. The amount of entrained gas by the stellar component is consistent with the mass lost by three very evolved stars (i.e., in the late asymptotic giant branch phase). However, we should keep in mind the limitations of the representation of the stellar-wind bubbles adopted in the present simulation, noted in Sect.~\ref{sijp}. Focusing on the entrained ISM/shear component, we observe that, once S3 has entered the jet spine ($t\geq 8000$~yr), the flux stabilizes at $1-2\times10^{-4}\,M_\odot{\rm \,yr^{-1}}$.

The right panel of Fig.~\ref{mload} shows the total mass flux along the jet spine ($r\leq 8.4\,{\rm pc}$) across the downstream outflow boundary, as a function of time. Initially, the mass flux across this section is $\sim  3.5\times10^{-4}\,M_\odot \, {\rm yr^{-1}}$, and it almost doubles once the stars have entered the jet spine, between 8 and 10~kyr, a relevant fraction of it being entrained ambient gas (see the left panel). If we compare both panels in Fig.~\ref{mload}, we see that the ISM/shear entrained material reaches up to 50\% of the jet mass flux during the simulation. Even though the limitations of our simulation make us to be cautious, this result clearly indicates that the scenario proposed is worth being taken into account and further studied.

\begin{figure}
\centering
\includegraphics[width=\linewidth]{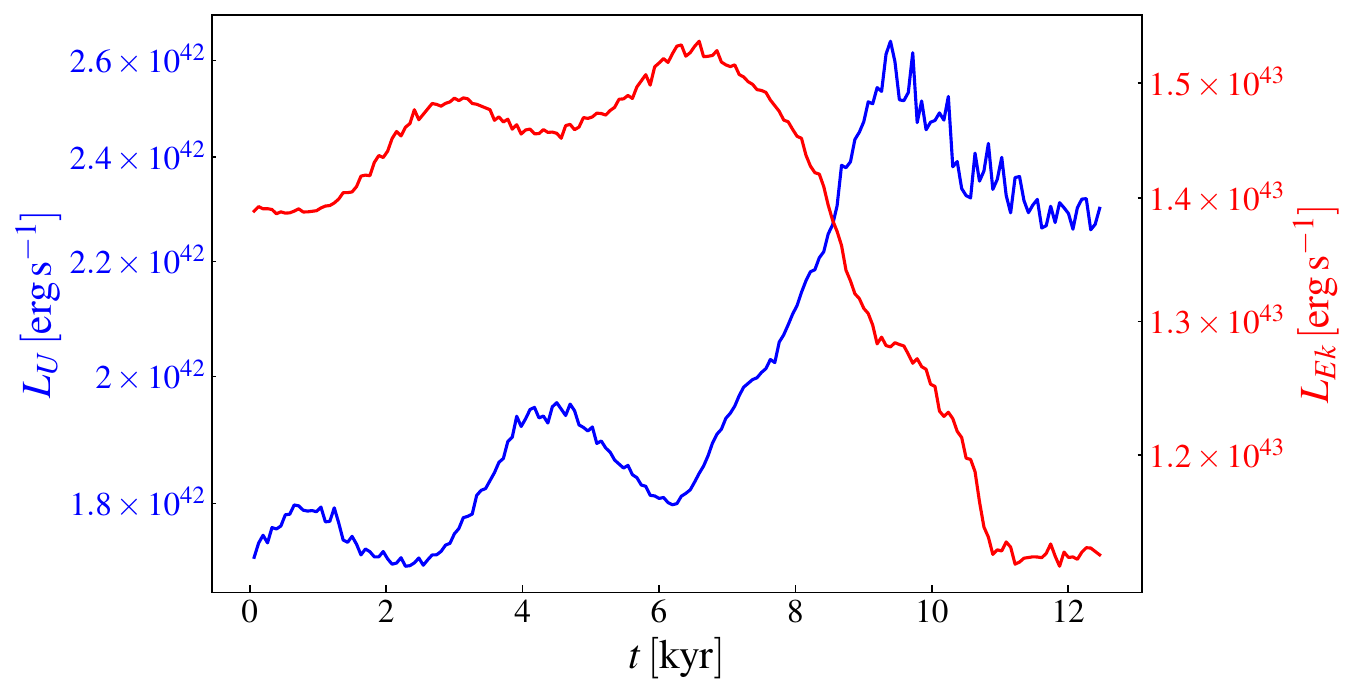}
\caption{Internal (blue line) and kinetic (red line) energy rates across the downstream boundary of the jet, within a radius of 8.4~pc. The total energy flux of the jet at injection, within this radius, is $L_{\rm j}=2.9\times 10^{43}\,\ergs$ (the fraction of $L_{\rm j}$ within the grid is half that value as we only simulate one half of the jet cross-section).}
\label{eflux}
\end{figure}

Figure~\ref{eflux} shows the jet energy flux, divided into internal and kinetic contributions,\footnote{$L_{U}= A \, \rho \, \Gamma \varepsilon \,W^{2}\,v_y$ is the internal energy flux, where $A$ is the cross-section, $\rho$ is the rest-mass density, $\varepsilon$ is the specific internal energy, $\Gamma$ is the adiabatic exponent, $W$ is the jet Lorentz factor, and $v_y$ is the axial velocity. $L_{\rm E_k} = A\, \rho\, W(W-1) \,c^{2} \, v_y$ is the kinetic energy flux.} at the downstream outflow boundary. Although the kinetic energy dominates the jet energy flux throughout the simulation run, we see a drop of $\geq 20\,$\% starting $7000\,$yr since the beginning of the simulation. At the same time, we see an increase of $\geq 30\,$\% in the internal energy flux, caused by the dissipation produced at shocks and turbulent mixing. The internal energy flux through the jet cross-section rises up to a $\sim 15-20\,$\% of the total jet power at injection in our grid within the jet spine (which is $L_{\rm j} \approx 1.4\times 10^{43}\,\ergs$). The amount of this energy that can be invested into accelerating non-thermal particles is uncertain, but it is the likely outcome of shocks, turbulence and shearing \citep{bosch12}. Therefore, the expected result is a general increase of the radiative output in the region, precisely as observed in FRI jets \citep{laing14}. 

%
%

\section{Discussion}\label{disc}

In this paper, we present a numerical experiment to probe the effect of stars crossing the jet surface and, in particular, their role as mediators in the entrainment of ISM gas, as proposed by \ptw. The simulation shows that the jet surface can be completely perturbed by the stellar crossing. Although this effect is probably exaggerated by the large size of the stellar bubbles in our simulation, the continuous interaction with many more smaller objects could also permanently induce a perturbation to the shear layer, as suggested by that work. Ideally, the scenario would require a large number of smaller stars (implying significantly larger resolution) impacting on a larger jet section, which is prohibitive in terms of the required resolution and computational resources. In addition, probing the effect of these perturbations far downstream of the interaction regions would require devoted simulations.

\subsection{Mass entrainment}

Given the aforementioned limitations intrinsic to this work, we focus on the plausibility and impact of the mass-load mechanisms introduced in Sect.~\ref{evol}. The result shown in Figure~\ref{mload} indeed supports the role of stars as an indirect tool to jet mass-loading. 

The relevance of the star-jet interaction process strongly depends on the number of large obstacles that penetrate the jet. In particular, stellar wind bubbles with $R_{\rm eq}\sim0.1$~pc like the ones used in our simulations require mass-loss rates of $\sim10^{-5}$--$10^{-4}\,$M$_\odot$~yr$^{-1}$ (accounting for uncertainties related to details of the wind-medium interaction), corresponding to stars in the late asymptotic giant branch (AGB) phase, which represent roughly one out of every $10^5$--$10^6$ stars. Considering the case of M87, the inner kiloparsec contains $\sim 10^{10}$~M$_\odot$ \citep{geb09}, and there could be $\sim10^4$--$10^5$ late AGB stars in the region, with $\sim10$-100 of them located inside the jet at any given time for a jet volume filling factor of $\sim10^{-3}$. Focusing on the VC effect, since it operates for at least $\sim10\%$ of the jet-crossing time in our simulations, this implies that, on average, roughly $\sim 1$--10 of these stars would be actively driving the VC mechanism near the jet boundary at any given moment. In contrast, the diversion of shear flow material is expected to act all through the shear-layer crossing, until the bow-shock tip is located inside the jet.

Less evolved stars, although associated with much smaller wind bubbles, are far more numerous and may contribute to mass entrainment through two complementary channels. Their direct contribution of both processes (shear flow diversion and VC) is expected to be much weaker because the associated entrainment scales approximately as $R_{\rm eq}^{3}$ (i.e., proportional to the affected boundary surface, which scales as $R_{\rm eq}^2$, times the interaction duration, which scales as $R_{\rm eq}$). However, their large numbers may compensate, at least partially, for the lower efficiency of each individual interaction. In addition, as discussed previously, these stars may seed perturbations that promote instability growth and mixing on larger scales (see next Section). In summary, the combined effect of a diverse population of obstacles spanning a range of sizes may also enhance ambient-gas entrainment beyond that produced by the largest bubbles alone. Moreover, the toroidal velocity expected to be present in jets can help to distribute the mixed material around the jet, homogenizing mass-load in a timescale $\sim R_j/v_\phi$. For $v_\phi\sim 10^{-3}c$ this implies timescales of $~10^4-10^5\,{\rm yr}$, and advection distances of a few kiloparsecs for moderately relativistic velocities, and even less for slowly moving material such as dense streams of entrained gas.

Regarding the entrainment from the stellar bubbles in our simulation, \pbb~ showed that the disruption time of shocked wind bubbles can be $\gtrsim 10^3\,{\rm yr}$. Given that our simulations extend for $\sim 10^4$~yr, one would expect the bubbles to be completely ablated by the end of the calculation. However, our numerical resolution is insufficient to properly describe the erosion of the bubbles and the much smaller scales at which equilibrium with the stellar wind would be established (as we said previously, due to our computational limitations the bubbles would be artificially disrupted within a few years). Thus, we artificially replenished them throughout the simulations, restoring their original properties after each iteration. The total mass estimated to be entrained in the jet from the bubbles is $0.75 M_\odot$ (see Sect.~\ref{ed}), although this contribution is only computed to isolate the entrainment of ISM/shear material produced by the star-jet interaction. We note that AGB star bubbles entering the jet could actually lead to a sudden release of mass akin to a somewhat under-massive supernova ejecta \citep{torres-alba19,lon25}.

%
\begin{figure*}
\centering
\includegraphics[width=0.48\linewidth]{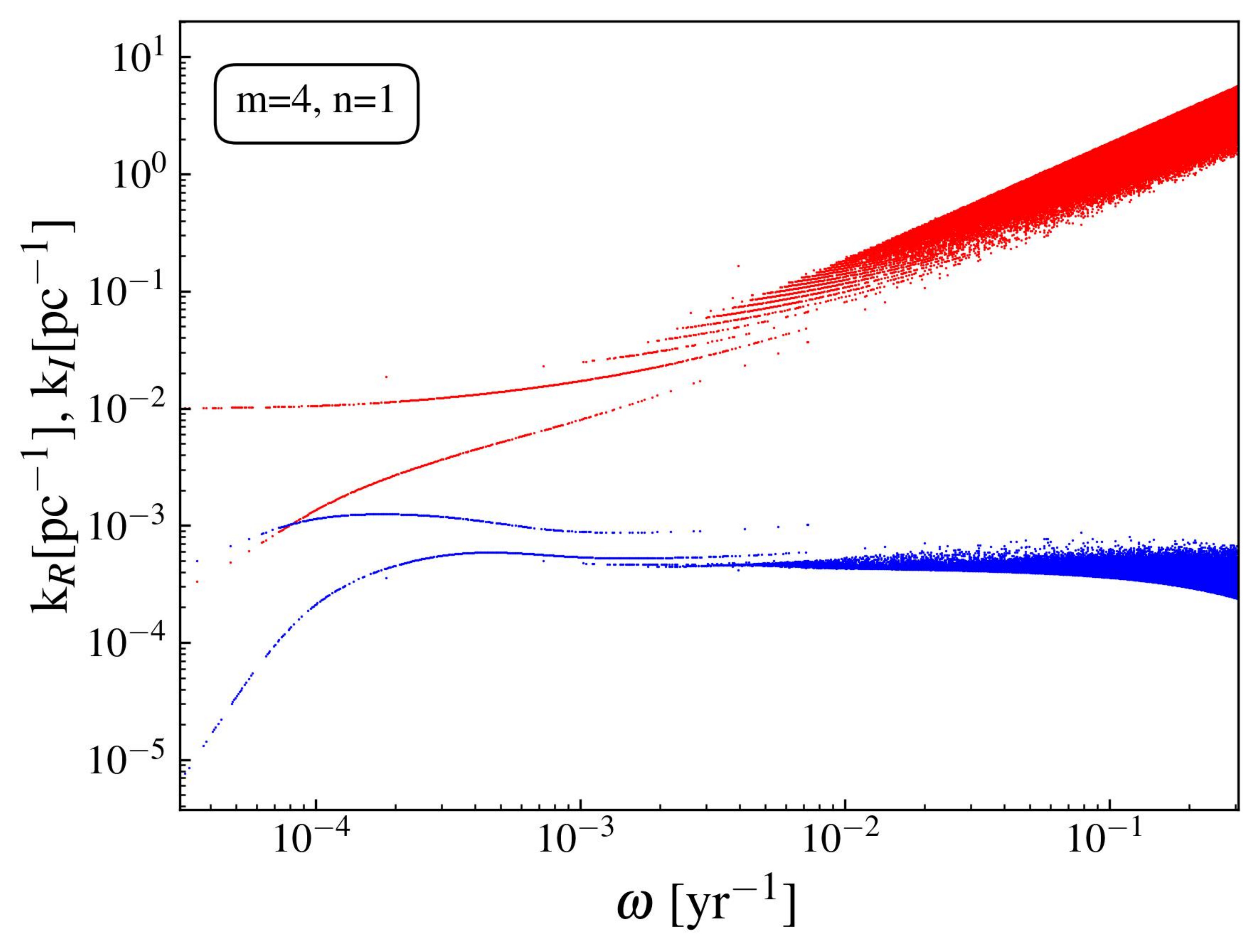}
\includegraphics[width=0.48\linewidth]{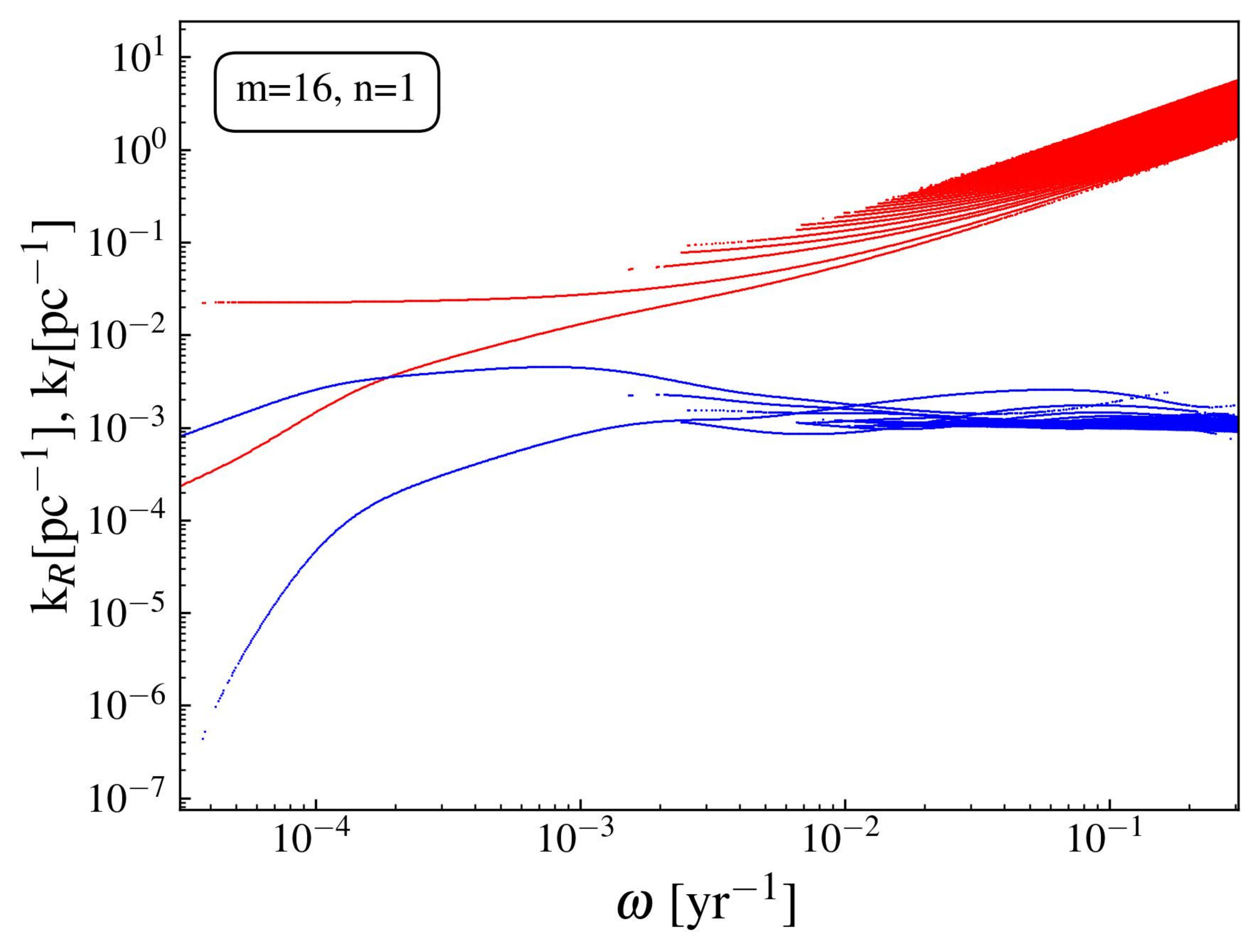} 
\caption{Linear problem solutions for the development of KH helical modes in the simulated system, in the spatial approach (real frequency and complex wavenumber) for different shear layer widths. The left panel shows the result for a wider shear layer than the right panel. The red lines indicate the real part of the wavenumber ($k_R$), whereas the blue lines indicate the imaginary part ($k_I$). The latter is the inverse of the growth length. Each of the red lines has its corresponding blue line, each one indicating the real and imaginary part of the solution. The discontinuous aspect of the lines is due to the point-by-point root finder that we use (see the text).}
\label{stab}
\end{figure*}
%

\subsection{On the development of instabilities}

Although our simulation does not allow us to study the development of instabilities and turbulent mixing up to scales of hundreds of kpc, we can perform a stability study to estimate the growth distances of the excited Kelvin-Helmholtz/shear instability modes either by impacts of small stars with the suggested frequency, $\sim 0.001-0.01\,{\rm yr^{-1}}$ (\ptw), or of larger stars as in the simulated scenario, $\sim 10^{-4}\,{\rm yr^{-1}}$. Figure~\ref{stab} shows the solutions found for the stability problem for the helical modes of the simulated jet, using two different shear-layer widths. We only include the helical pattern because the impact of a star creates an asymmetric perturbation. The transition from the jet to the ambient is established via the expression $x=x_0/\cosh{r^m}$, and the width of the layer $\Delta$ is thus parametrised using the value of the exponent $m$ (for the values we use, 4 and 16, this implies $\Delta \sim 0.90$ and $0.22\,R_j$, respectively). The linearized stability equation is solved using a shooting method that we apply from the jet axis to its environment, as explained in \citet{2007PhRvE..75e6312P,2007A&A...469L..23P,2019A&A...627A..79V}. This is a point-by-point root finder and convergence is sometimes difficult or computationally demanding to achieve, which results in the point-like aspect of the solutions. However, the plots are enough to assess the expected wavelengths and growth lengths produced by the impact of stars with the indicated frequencies. 

The plots show that the largest values of $k_I$ (i.e., the shortest growth lengths) correspond to long-wavelength perturbations, with both wavelengths and growth lengths ranging from hundreds to thousands of parsecs. However, observations favour much shorter wavelengths, and the linear growth-lengths become less relevant if stellar impacts already drive the perturbations into the non-linear regime.\footnote{$k_I$ gives the inverse of the distance in which the amplitude of the perturbation grows by an exponential factor in the linear regime. If the amplitude of the perturbation is already large (non-linear), it becomes irrelevant.} Within the frequency range of interest, there exist short-wavelength modes ($\sim$ 10~pc; $k_R \gtrsim 10^{-1}\,{\rm pc}^{-1}$) with very long growth lengths (low $k_I$). These modes are mainly associated with the highest frequencies considered, which correspond to impacts by less-evolved stars. These stars could be then the main triggers of short wavelength instabilities in the jet, and thus of an instability-related mass-load. The effect of the shear layer is reflected in slightly lower values of $k_I$ for the wider layer ($m=4$), although this has little impact, as discussed above, on the scenario described, which thus remains plausible regardless of the shear-layer width.

The penetration of stars in jets can also trigger the development of Rayleigh-Taylor or centrifugal instabilities (RTI and CFI, respectively). Regarding RTI,
\citet{matsumoto17} derived a stability criterion that requires $W_{\rm j}^2 \rho_{\rm j} h'_{\rm j} > W_{\rm a}^2 \rho_{\rm a} h'_{\rm a}$ (with $h'= 1 + \Gamma^2 p/\left[(1+\Gamma) \rho\right]$; note this is not the specific enthalpy)
as the condition for instability development. In our case $W_{\rm j}^2 \rho_{\rm j} h'_{\rm j} \ll W_{\rm a}^2 \rho_{\rm a} h'_{\rm a}$, which excludes this possibility. Taking into account that $W_{\rm a}=1$ and $h'_{\rm a}=1$, $W_{\rm a}^2 \rho_{\rm a} h'_{\rm a}$ depends only on $\rho_a$. For the chosen value, $W_{\rm j}^2 \rho_{\rm j} h'_{\rm j} $ should be two orders of magnitude larger than the one used to make the system unstable to RTI \citep[see also][]{2019MNRAS.490.4271M}.

In the case of the CFI, the stability criterion is based on a similar expression as that for RTI, involving the azimuthal velocity \citep[$\rho h W^2 (\Omega R^2)^2$, where $\Omega$ is the angular velocity and $R$ is the cylindrical radius;][]{gourg18b}. In \citet{gourg18a}, the authors considered the radial acceleration produced by expansion and recollimation (with an effective acceleration pointing towards the jet axis) to study the development of this instability in FRI jets. Later, \citet{komissarov19} showed that RTI is suppressed except in the case of jets with low magnetization, but this is precisely expected at tens to hundred of parsecs after jet acceleration has taken place, on the one hand, and after mass entrainment has already started \citep{castillo2021}.

An argument that has been used as a caveat for RTI to be the cause of jet deceleration is that, following \citet{gourg18a}, a strong recollimation shock is expected in this scenario, whereas such shock are not observed in canonical FRI jets \citep{laing14}. However, if entrainment occurs efficiently due to both the star-driven entrainment and the development of the RTI, the jet could become transonic, thus avoiding the formation of such a shock, which gives a plausible scenario for FRI jet deceleration. If the stars can act as initial perturbation mechanisms to couple to the RTI modes, the distances in which the RTI instability could develop in an expanding jet remain to be estimated from linear and/or post-linear analysis.

\section{Summary and conclusions} \label{sum}

We have run a numerical experiment to study the role of star-jet interaction as a driver of ISM gas entrainment in relativistic jets in AGN. For this, we have set up three stellar bubbles characterized as the shocked wind stellar regions, and allow them to move into the jet at different times. Our results show different interesting effects, such as the shielding of downstream stars by upstream bow shocks or the enhanced temporal propagation of mixing layers towards the jet axis as the stars cross the jet surface. We have then studied the contribution of ISM and shear flow to the entrainment of the jet during this process. We have shown that, even in this limited experiment, piercing of the jet surface by stars can act as a trigger of jet deceleration and energy dissipation via the star-driven (both through the bow-shock diversion of shear material and the VC effect) and direct mass-load from the strongest mass-loss stars on smaller scales, and instability growth on larger ones.

The model presented in \ptw\, suggested that stars going in and out of jets could trigger the development of small-scale instabilities and the formation of a mixing layer that would develop downstream, from the jet boundary to its axis. We have shown that stars can indeed strongly perturb the jet surface as they cross it. However, the development of instabilities is expected to require larger scales ($\sim 10-100\,{\rm pc}$) than those covered by our simulations. Nevertheless, we have  also shown that, in addition to that possible effect, the stellar motion creates a diversion of shear material into the jet, plus a low-pressure wake behind the star, which can act as an attractor for shocked ambient material into the jet (VC mechanism). In this scenario, entrainment takes place through shocks and turbulent mixing \citep[see Figs.~\ref{8snaprho}-\ref{8snaprhot}, which also favours the necessary dissipation to explain jet brightening in the decelerating region][]{laing14}. According to our results, dissipation can be efficient, consuming up to 20\% of the kinetic energy in our simulations (i.e., even for only three stars).

The reason why we expect this process to contribute to jet deceleration in FRI jets implicitly assumes it will not be so for FRII jets, and this requires explanation. First, this expectation is based on previous numerical simulations where mass-load from stellar winds was added to jets with different powers and/or properties \citep[e.g.,][]{perucho14,angcast21} and the results showed that the effect is smaller and tends to be negligible for powers above $10^{44}\,{\rm erg/s}$. Second, in \ptw, the author also pointed out that the entrainment by the stars crossing the jet is weakened if the shocked gas bubble is destroyed at the shear layer (see Sect. 3.5 in that paper), as it may well be in powerful FRIIs, which possibly have shielding winds protecting the jet spine \citep{boccardi21}. It is worth mentioning that the impact with large stars can, at the least, involve relevant transient entrainment events. Whether it can contribute to jet deceleration in the long term will need to be tested by numerical simulations involving larger scales and, when possible, resolving properly a varied and realistic stellar population.

Future work should tackle the coupling of the perturbations produced by stars on the jet surface to instabilities, the possible generation of a mixing layer in distance, as proposed by \ptw, and whether this is compatible with the enhancement of jet brightness along the decelerating length noted in \citet{laing14}, driven by shocks and turbulent mixing. We currently design new simulations for magnetized jets, with the aim to study whether the jet magnetic fields can play a role in the process and affect our conclusions. 

Finally, it is worth noting that the strong and sudden energy reprocessing involved in the jet-star interaction, in which large amounts of mass are entrained by the jet in a relatively small jet region, are likely to lead to significant non-thermal activity via Fermi-type mechanisms such as diffusive shock, stochastic and shear acceleration \citep[e.g.][]{rie07}. This may be relevant regarding both potential non-thermal emission signatures of the process in different energy bands, and cosmic ray acceleration, to be studied in future work.

\begin{acknowledgements}
This work has received financial support from the Spanish Ministry of Science and Innovation under grants 
PID2022-136828NB-C41/AEI/10.13039/501100011033/ERDF/EU and PID2022-136828NB-C43/AEI/10.13039/501100011033/ERDF/EU, from the Generalitat Valenciana through grant \texttt{CIPROM/2022/49}, and the María de Maeztu award to the ICCUB CEX2024-001451-M. V.B-R. is Correspondent Researcher of CONICET, Argentina, at the IAR.
\end{acknowledgements}

\bibliographystyle{aa}
\bibliography{biblio}

@ARTICLE{lon25,
       author = {{Longo}, B. and {Perucho}, M. and {Bosch-Ramon}, V. and {Mart{\'\i}}, J.~M. and {Fichet de Clairfontaine}, G.},
        title = "{Relativistic hydrodynamics simulations of supernova explosions within extragalactic jets}",
      journal = {\aap},
         year = 2025,
        month = dec,
       volume = {704},
          eid = {A172},
        pages = {A172},
          doi = {10.1051/0004-6361/202555849},
archivePrefix = {arXiv},
       eprint = {2510.04570},
 primaryClass = {astro-ph.HE},
       adsurl = {https://ui.adsabs.harvard.edu/abs/2025A&A...704A.172L}
}

@ARTICLE{rie07,
       author = {{Rieger}, Frank M. and {Bosch-Ramon}, Valent{\'\i} and {Duffy}, Peter},
        title = "{Fermi acceleration in astrophysical jets}",
      journal = {\apss},
         year = 2007,
        month = jun,
       volume = {309},
       number = {1-4},
        pages = {119-125},
          doi = {10.1007/s10509-007-9466-z},
archivePrefix = {arXiv},
       eprint = {astro-ph/0610141},
 primaryClass = {astro-ph},
       adsurl = {https://ui.adsabs.harvard.edu/abs/2007Ap&SS.309..119R}
}

@ARTICLE{geb09,
       author = {{Gebhardt}, Karl and {Thomas}, Jens},
        title = "{The Black Hole Mass, Stellar Mass-to-Light Ratio, and Dark Halo in M87}",
      journal = {\apj},
         year = 2009,
        month = aug,
       volume = {700},
       number = {2},
        pages = {1690-1701},
          doi = {10.1088/0004-637X/700/2/1690},
archivePrefix = {arXiv},
       eprint = {0906.1492},
 primaryClass = {astro-ph.CO},
       adsurl = {https://ui.adsabs.harvard.edu/abs/2009ApJ...700.1690G}
}

@ARTICLE{bicknell86,
       author = {{Bicknell}, G.~V.},
        title = "{A Model for the Surface Brightness of a Turbulent, Low Mach Number Jet. II. The Global Energy Budget and Radiative Losses}",
      journal = {\apj},
         year = 1986,
        month = jan,
       volume = {300},
        pages = {591},
          doi = {10.1086/163836},
       adsurl = {https://ui.adsabs.harvard.edu/abs/1986ApJ...300..591B}
}

@ARTICLE{bicknell84,
       author = {{Bicknell}, G.~V.},
        title = "{A model for the surface brightness of a turbulent low mach number jet. I. Theoretical development and application to 3C 31.}",
      journal = {\apj},
         year = 1984,
        month = nov,
       volume = {286},
        pages = {68-87},
          doi = {10.1086/162577},
       adsurl = {https://ui.adsabs.harvard.edu/abs/1984ApJ...286...68B}
}

@ARTICLE{blandford77,
       author = {{Blandford}, R.~D. and {Znajek}, R.~L.},
        title = "{Electromagnetic extraction of energy from Kerr black holes.}",
      journal = {\mnras},
         year = 1977,
        month = may,
       volume = {179},
        pages = {433-456},
          doi = {10.1093/mnras/179.3.433},
       adsurl = {https://ui.adsabs.harvard.edu/abs/1977MNRAS.179..433B}
}

@ARTICLE{fr74,
       author = {{Fanaroff}, B.~L. and {Riley}, J.~M.},
        title = "{The morphology of extragalactic radio sources of high and low luminosity}",
      journal = {\mnras},
         year = 1974,
        month = may,
       volume = {167},
        pages = {31P-36P},
          doi = {10.1093/mnras/167.1.31P},
       adsurl = {https://ui.adsabs.harvard.edu/abs/1974MNRAS.167P..31F}
}

@ARTICLE{fichet25,
       author = {{Fichet de Clairfontaine}, G. and {Perucho}, M. and {Mart{\'\i}}, J.~M. and {Kovalev}, Y.~Y.},
        title = "{Dynamic and radiative implications of jet{\textendash}star interactions in AGN jets}",
      journal = {\aap},
         year = 2025,
        month = jan,
       volume = {693},
          eid = {A270},
        pages = {A270},
          doi = {10.1051/0004-6361/202451914},
archivePrefix = {arXiv},
       eprint = {2412.07945},
 primaryClass = {astro-ph.HE},
       adsurl = {https://ui.adsabs.harvard.edu/abs/2025A&A...693A.270F}
}

@ARTICLE{hub06,
       author = {{Hubbard}, A. and {Blackman}, E.~G.},
        title = "{Active galactic nuclei jet mass loading and truncation by stellar winds}",
      journal = {\mnras},
         year = 2006,
        month = oct,
       volume = {371},
       number = {4},
        pages = {1717-1721},
          doi = {10.1111/j.1365-2966.2006.10808.x},
archivePrefix = {arXiv},
       eprint = {astro-ph/0604585},
 primaryClass = {astro-ph},
       adsurl = {https://ui.adsabs.harvard.edu/abs/2006MNRAS.371.1717H}
}

@ARTICLE{torres-alba19,
       author = {{Torres-Alb{\`a}}, N{\'u}ria and {Bosch-Ramon}, Valent{\'\i}},
        title = "{Gamma rays from red giant wind bubbles entering the jets of elliptical host blazars}",
      journal = {\aap},
         year = 2019,
        month = mar,
       volume = {623},
          eid = {A91},
        pages = {A91},
          doi = {10.1051/0004-6361/201833697},
archivePrefix = {arXiv},
       eprint = {1902.05008},
 primaryClass = {astro-ph.HE},
       adsurl = {https://ui.adsabs.harvard.edu/abs/2019A&A...623A..91T}
}

@ARTICLE{araudo13,
       author = {{Araudo}, A.~T. and {Bosch-Ramon}, V. and {Romero}, G.~E.},
        title = "{Gamma-ray emission from massive stars interacting with active galactic nuclei jets}",
      journal = {\mnras},
         year = 2013,
        month = dec,
       volume = {436},
       number = {4},
        pages = {3626-3639},
          doi = {10.1093/mnras/stt1840},
archivePrefix = {arXiv},
       eprint = {1309.7114},
 primaryClass = {astro-ph.HE},
       adsurl = {https://ui.adsabs.harvard.edu/abs/2013MNRAS.436.3626A}
}

@ARTICLE{komissarov94,
       author = {{Komissarov}, S.~S.},
        title = "{Mass-Loaded Relativistic Jets}",
      journal = {\mnras},
         year = 1994,
        month = jul,
       volume = {269},
        pages = {394},
          doi = {10.1093/mnras/269.2.394},
       adsurl = {https://ui.adsabs.harvard.edu/abs/1994MNRAS.269..394K}
}

@ARTICLE{perucho20,
       author = {{Perucho}, Manel},
        title = "{Triggering mixing and deceleration in FRI jets: a solution}",
      journal = {\mnras},
         year = 2020,
        month = may,
       volume = {494},
       number = {1},
        pages = {L22-L26},
          doi = {10.1093/mnrasl/slaa031},
archivePrefix = {arXiv},
       eprint = {2002.05434},
 primaryClass = {astro-ph.HE},
       adsurl = {https://ui.adsabs.harvard.edu/abs/2020MNRAS.494L..22P}
}

@ARTICLE{perucho14,
       author = {{Perucho}, M. and {Mart{\'\i}}, J.~M. and {Laing}, R.~A. and {Hardee}, P.~E.},
        title = "{On the deceleration of Fanaroff-Riley Class I jets: mass loading by stellar winds}",
      journal = {\mnras},
         year = 2014,
        month = jun,
       volume = {441},
       number = {2},
        pages = {1488-1503},
          doi = {10.1093/mnras/stu676},
archivePrefix = {arXiv},
       eprint = {1404.1209},
 primaryClass = {astro-ph.HE},
       adsurl = {https://ui.adsabs.harvard.edu/abs/2014MNRAS.441.1488P}
}

@ARTICLE{boccardi21,
       author = {{Boccardi}, B. and {Perucho}, M. and {Casadio}, C. and {Grandi}, P. and {Macconi}, D. and {Torresi}, E. and {Pellegrini}, S. and {Krichbaum}, T.~P. and {Kadler}, M. and {Giovannini}, G. and {Karamanavis}, V. and {Ricci}, L. and {Madika}, E. and {Bach}, U. and {Ros}, E. and {Giroletti}, M. and {Zensus}, J.~A.},
        title = "{Jet collimation in NGC 315 and other nearby AGN}",
      journal = {\aap},
         year = 2021,
        month = mar,
       volume = {647},
          eid = {A67},
        pages = {A67},
          doi = {10.1051/0004-6361/202039612},
archivePrefix = {arXiv},
       eprint = {2012.14831},
 primaryClass = {astro-ph.HE},
       adsurl = {https://ui.adsabs.harvard.edu/abs/2021A&A...647A..67B}
}

@ARTICLE{wykes15,
       author = {{Wykes}, Sarka and {Hardcastle}, Martin J. and {Karakas}, Amanda I. and {Vink}, Jorick S.},
        title = "{Internal entrainment and the origin of jet-related broad-band emission in Centaurus A}",
      journal = {\mnras},
         year = 2015,
        month = feb,
       volume = {447},
       number = {1},
        pages = {1001-1013},
          doi = {10.1093/mnras/stu2440},
archivePrefix = {arXiv},
       eprint = {1409.5785},
 primaryClass = {astro-ph.HE},
       adsurl = {https://ui.adsabs.harvard.edu/abs/2015MNRAS.447.1001W}
}

@ARTICLE{laing14,
       author = {{Laing}, R.~A. and {Bridle}, A.~H.},
        title = "{Systematic properties of decelerating relativistic jets in low-luminosity radio galaxies}",
      journal = {\mnras},
         year = 2014,
        month = feb,
       volume = {437},
       number = {4},
        pages = {3405-3441},
          doi = {10.1093/mnras/stt2138},
archivePrefix = {arXiv},
       eprint = {1311.1015},
 primaryClass = {astro-ph.CO},
       adsurl = {https://ui.adsabs.harvard.edu/abs/2014MNRAS.437.3405L}
}

@ARTICLE{castillo2021,
       author = {{Angl{\'e}s-Castillo}, Andreu and {Perucho}, Manel and {Mart{\'\i}}, Jos{\'e} Mar{\'\i}a and {Laing}, Robert A.},
        title = "{On the deceleration of Fanaroff-Riley Class I jets: mass loading of magnetized jets by stellar winds}",
      journal = {\mnras},
         year = 2021,
        month = jan,
       volume = {500},
       number = {1},
        pages = {1512-1530},
          doi = {10.1093/mnras/staa3291},
archivePrefix = {arXiv},
       eprint = {2010.10234},
 primaryClass = {astro-ph.HE},
       adsurl = {https://ui.adsabs.harvard.edu/abs/2021MNRAS.500.1512A}
}

@ARTICLE{perucho17,
       author = {{Perucho}, M. and {Bosch-Ramon}, V. and {Barkov}, M.~V.},
        title = "{Impact of red giant/AGB winds on active galactic nucleus jet propagation}",
      journal = {\aap},
         year = 2017,
        month = oct,
       volume = {606},
          eid = {A40},
        pages = {A40},
          doi = {10.1051/0004-6361/201630117},
archivePrefix = {arXiv},
       eprint = {1706.06301},
 primaryClass = {astro-ph.HE},
       adsurl = {https://ui.adsabs.harvard.edu/abs/2017A&A...606A..40P}
}

@ARTICLE{bosch12,
       author = {{Bosch-Ramon}, V. and {Perucho}, M. and {Barkov}, M.~V.},
        title = "{Clouds and red giants interacting with the base of AGN jets.}",
      journal = {\aap},
         year = 2012,
        month = mar,
       volume = {539},
          eid = {A69},
        pages = {A69},
          doi = {10.1051/0004-6361/201118622},
archivePrefix = {arXiv},
       eprint = {1201.5279},
 primaryClass = {astro-ph.HE},
       adsurl = {https://ui.adsabs.harvard.edu/abs/2012A&A...539A..69B}
}

@ARTICLE{bowman96,
       author = {{Bowman}, M. and {Leahy}, J.~P. and {Komissarov}, S.~S.},
        title = "{The deceleration of relativistic jets by entrainment}",
      journal = {\mnras},
         year = 1996,
        month = apr,
       volume = {279},
        pages = {899},
          doi = {10.1093/mnras/279.3.899},
       adsurl = {https://ui.adsabs.harvard.edu/abs/1996MNRAS.279..899B}
}

@ARTICLE{angcast21,
       author = {{Angl{\'e}s-Castillo}, Andreu and {Perucho}, Manel and {Mart{\'\i}}, Jos{\'e} Mar{\'\i}a and {Laing}, Robert A.},
        title = "{On the deceleration of Fanaroff-Riley Class I jets: mass loading of magnetized jets by stellar winds}",
      journal = {\mnras},
         year = 2021,
        month = jan,
       volume = {500},
       number = {1},
        pages = {1512-1530},
          doi = {10.1093/mnras/staa3291},
archivePrefix = {arXiv},
       eprint = {2010.10234},
 primaryClass = {astro-ph.HE},
       adsurl = {https://ui.adsabs.harvard.edu/abs/2021MNRAS.500.1512A}
}

@ARTICLE{matsumoto17,
       author = {{Matsumoto}, Jin and {Aloy}, Miguel A. and {Perucho}, Manel},
        title = "{Linear theory of the Rayleigh-Taylor instability at a discontinuous surface of a relativistic flow}",
      journal = {\mnras},
         year = 2017,
        month = dec,
       volume = {472},
       number = {2},
        pages = {1421-1431},
          doi = {10.1093/mnras/stx2012},
archivePrefix = {arXiv},
       eprint = {1707.04706},
 primaryClass = {astro-ph.HE},
       adsurl = {https://ui.adsabs.harvard.edu/abs/2017MNRAS.472.1421M}
}

@ARTICLE{gourg18b,
       author = {{Gourgouliatos}, Konstantinos N. and {Komissarov}, Serguei S.},
        title = "{Relativistic centrifugal instability}",
      journal = {\mnras},
         year = 2018,
        month = mar,
       volume = {475},
       number = {1},
        pages = {L125-L129},
          doi = {10.1093/mnrasl/sly016},
archivePrefix = {arXiv},
       eprint = {1710.01345},
 primaryClass = {astro-ph.HE},
       adsurl = {https://ui.adsabs.harvard.edu/abs/2018MNRAS.475L.125G}
}

@ARTICLE{gourg18a,
       author = {{Gourgouliatos}, Konstantinos N. and {Komissarov}, Serguei S.},
        title = "{Reconfinement and loss of stability in jets from active galactic nuclei}",
      journal = {Nature Astronomy},
         year = 2018,
        month = dec,
       volume = {2},
        pages = {167-171},
          doi = {10.1038/s41550-017-0338-3},
archivePrefix = {arXiv},
       eprint = {1806.05683},
 primaryClass = {astro-ph.HE},
       adsurl = {https://ui.adsabs.harvard.edu/abs/2018NatAs...2..167G}
}

@ARTICLE{komissarov19,
       author = {{Komissarov}, Serguei S. and {Gourgouliatos}, Konstantinos N. and {Matsumoto}, Jin},
        title = "{Magnetic Inhibition of Centrifugal Instability in Astrophysical Jets}",
      journal = {arXiv e-prints},
         year = 2019,
        month = may,
          eid = {arXiv:1905.11650},
        pages = {arXiv:1905.11650},
          doi = {10.48550/arXiv.1905.11650},
archivePrefix = {arXiv},
       eprint = {1905.11650},
 primaryClass = {astro-ph.HE},
       adsurl = {https://ui.adsabs.harvard.edu/abs/2019arXiv190511650K}
}

@article{perucho07,
    author = {Perucho, M. and Martí, J. M.},
    title = {A numerical simulation of the evolution and fate of a Fanaroff–Riley type I jet. The case of 3C 31},
    journal = {Monthly Notices of the Royal Astronomical Society},
    volume = {382},
    number = {2},
    pages = {526-542},
    year = {2007},
    month = {11},
    issn = {0035-8711},
    doi = {10.1111/j.1365-2966.2007.12454.x},
    url = {https://doi.org/10.1111/j.1365-2966.2007.12454.x},
    eprint = {https://academic.oup.com/mnras/article-pdf/382/2/526/3411374/mnras0382-0526.pdf},
}

@Article{perucho19,
AUTHOR = {Perucho, Manel},
TITLE = {Dissipative Processes and Their Role in the Evolution of Radio Galaxies},
JOURNAL = {Galaxies},
VOLUME = {7},
YEAR = {2019},
NUMBER = {3},
ARTICLE-NUMBER = {70},
URL = {https://www.mdpi.com/2075-4434/7/3/70},
ISSN = {2075-4434},
DOI = {10.3390/galaxies7030070}
}

@ARTICLE{perucho10,
       author = {{Perucho}, M. and {Mart{\'\i}}, J.~M. and {Cela}, J.~M. and {Hanasz}, M. and {de La Cruz}, R. and {Rubio}, F.},
        title = "{Stability of three-dimensional relativistic jets: implications for jet collimation}",
      journal = {\aap},
         year = 2010,
        month = sep,
       volume = {519},
          eid = {A41},
        pages = {A41},
          doi = {10.1051/0004-6361/200913012},
archivePrefix = {arXiv},
       eprint = {1005.4332},
 primaryClass = {astro-ph.HE},
       adsurl = {https://ui.adsabs.harvard.edu/abs/2010A&A...519A..41P}
}

@ARTICLE{synge57,
author={{Synge}, J.L},
title="{The Relativistic Gas}",
year = 1957,
}

@ARTICLE{2020A&A...633L...1R,
       author = {{Ros}, E. and {Kadler}, M. and {Perucho}, M. and {Boccardi}, B. and {Cao}, H. -M. and {Giroletti}, M. and {Krau{\ss}}, F. and {Ojha}, R.},
        title = "{Apparent superluminal core expansion and limb brightening in the candidate neutrino blazar TXS 0506+056}",
      journal = {\aap},
         year = 2020,
        month = jan,
       volume = {633},
          eid = {L1},
        pages = {L1},
          doi = {10.1051/0004-6361/201937206},
archivePrefix = {arXiv},
       eprint = {1912.01743},
 primaryClass = {astro-ph.GA},
       adsurl = {https://ui.adsabs.harvard.edu/abs/2020A&A...633L...1R}
}

@ARTICLE{mistele24,
       author = {{Mistele}, Tobias and {McGaugh}, Stacy and {Lelli}, Federico and {Schombert}, James and {Li}, Pengfei},
        title = "{Indefinitely Flat Circular Velocities and the Baryonic Tully{\textendash}Fisher Relation from Weak Lensing}",
      journal = {\apjl},
         year = 2024,
        month = jul,
       volume = {969},
       number = {1},
          eid = {L3},
        pages = {L3},
          doi = {10.3847/2041-8213/ad54b0},
archivePrefix = {arXiv},
       eprint = {2406.09685},
 primaryClass = {astro-ph.GA},
       adsurl = {https://ui.adsabs.harvard.edu/abs/2024ApJ...969L...3M}
}

@ARTICLE{2001ApJ...552..508G,
       author = {{Giovannini}, G. and {Cotton}, W.~D. and {Feretti}, L. and {Lara}, L. and {Venturi}, T.},
        title = "{VLBI Observations of a Complete Sample of Radio Galaxies: 10 Years Later}",
      journal = {\apj},
         year = 2001,
        month = may,
       volume = {552},
       number = {2},
        pages = {508-526},
          doi = {10.1086/320581},
archivePrefix = {arXiv},
       eprint = {astro-ph/0101096},
 primaryClass = {astro-ph},
       adsurl = {https://ui.adsabs.harvard.edu/abs/2001ApJ...552..508G}
}

@ARTICLE{2019ApJ...875L...1E,
       author = {{Event Horizon Telescope Collaboration} and {Akiyama}, Kazunori and {Alberdi}, Antxon and {Alef}, Walter and {Asada}, Keiichi and {Azulay}, Rebecca and {Baczko}, Anne-Kathrin and {Ball}, David and {Balokovi{\'c}}, Mislav and {Barrett}, John and {Bintley}, Dan and {Blackburn}, Lindy and {Boland}, Wilfred and {Bouman}, Katherine L. and {Bower}, Geoffrey C. and {Bremer}, Michael and {Brinkerink}, Christiaan D. and {Brissenden}, Roger and {Britzen}, Silke and {Broderick}, Avery E. and {Broguiere}, Dominique and {Bronzwaer}, Thomas and {Byun}, Do-Young and {Carlstrom}, John E. and {Chael}, Andrew and {Chan}, Chi-kwan and {Chatterjee}, Shami and {Chatterjee}, Koushik and {Chen}, Ming-Tang and {Chen}, Yongjun and {Cho}, Ilje and {Christian}, Pierre and {Conway}, John E. and {Cordes}, James M. and {Crew}, Geoffrey B. and {Cui}, Yuzhu and {Davelaar}, Jordy and {De Laurentis}, Mariafelicia and {Deane}, Roger and {Dempsey}, Jessica and {Desvignes}, Gregory and {Dexter}, Jason and {Doeleman}, Sheperd S. and {Eatough}, Ralph P. and {Falcke}, Heino and {Fish}, Vincent L. and {Fomalont}, Ed and {Fraga-Encinas}, Raquel and {Freeman}, William T. and {Friberg}, Per and {Fromm}, Christian M. and {G{\'o}mez}, Jos{\'e} L. and {Galison}, Peter and {Gammie}, Charles F. and {Garc{\'\i}a}, Roberto and {Gentaz}, Olivier and {Georgiev}, Boris and {Goddi}, Ciriaco and {Gold}, Roman and {Gu}, Minfeng and {Gurwell}, Mark and {Hada}, Kazuhiro and {Hecht}, Michael H. and {Hesper}, Ronald and {Ho}, Luis C. and {Ho}, Paul and {Honma}, Mareki and {Huang}, Chih-Wei L. and {Huang}, Lei and {Hughes}, David H. and {Ikeda}, Shiro and {Inoue}, Makoto and {Issaoun}, Sara and {James}, David J. and {Jannuzi}, Buell T. and {Janssen}, Michael and {Jeter}, Britton and {Jiang}, Wu and {Johnson}, Michael D. and {Jorstad}, Svetlana and {Jung}, Taehyun and {Karami}, Mansour and {Karuppusamy}, Ramesh and {Kawashima}, Tomohisa and {Keating}, Garrett K. and {Kettenis}, Mark and {Kim}, Jae-Young and {Kim}, Junhan and {Kim}, Jongsoo and {Kino}, Motoki and {Koay}, Jun Yi and {Koch}, Patrick M. and {Koyama}, Shoko and {Kramer}, Michael and {Kramer}, Carsten and {Krichbaum}, Thomas P. and {Kuo}, Cheng-Yu and {Lauer}, Tod R. and {Lee}, Sang-Sung and {Li}, Yan-Rong and {Li}, Zhiyuan and {Lindqvist}, Michael and {Liu}, Kuo and {Liuzzo}, Elisabetta and {Lo}, Wen-Ping and {Lobanov}, Andrei P. and {Loinard}, Laurent and {Lonsdale}, Colin and {Lu}, Ru-Sen and {MacDonald}, Nicholas R. and {Mao}, Jirong and {Markoff}, Sera and {Marrone}, Daniel P. and {Marscher}, Alan P. and {Mart{\'\i}-Vidal}, Iv{\'a}n and {Matsushita}, Satoki and {Matthews}, Lynn D. and {Medeiros}, Lia and {Menten}, Karl M. and {Mizuno}, Yosuke and {Mizuno}, Izumi and {Moran}, James M. and {Moriyama}, Kotaro and {Moscibrodzka}, Monika and {M{\"u}ller}, Cornelia and {Nagai}, Hiroshi and {Nagar}, Neil M. and {Nakamura}, Masanori and {Narayan}, Ramesh and {Narayanan}, Gopal and {Natarajan}, Iniyan and {Neri}, Roberto and {Ni}, Chunchong and {Noutsos}, Aristeidis and {Okino}, Hiroki and {Olivares}, H{\'e}ctor and {Ortiz-Le{\'o}n}, Gisela N. and {Oyama}, Tomoaki and {{\"O}zel}, Feryal and {Palumbo}, Daniel C.~M. and {Patel}, Nimesh and {Pen}, Ue-Li and {Pesce}, Dominic W. and {Pi{\'e}tu}, Vincent and {Plambeck}, Richard and {PopStefanija}, Aleksandar and {Porth}, Oliver and {Prather}, Ben and {Preciado-L{\'o}pez}, Jorge A. and {Psaltis}, Dimitrios and {Pu}, Hung-Yi and {Ramakrishnan}, Venkatessh and {Rao}, Ramprasad and {Rawlings}, Mark G. and {Raymond}, Alexander W. and {Rezzolla}, Luciano and {Ripperda}, Bart and {Roelofs}, Freek and {Rogers}, Alan and {Ros}, Eduardo and {Rose}, Mel and {Roshanineshat}, Arash and {Rottmann}, Helge and {Roy}, Alan L. and {Ruszczyk}, Chet and {Ryan}, Benjamin R. and {Rygl}, Kazi L.~J. and {S{\'a}nchez}, Salvador and {S{\'a}nchez-Arguelles}, David and {Sasada}, Mahito and {Savolainen}, Tuomas and {Schloerb}, F. Peter and {Schuster}, Karl-Friedrich and {Shao}, Lijing and {Shen}, Zhiqiang and {Small}, Des and {Sohn}, Bong Won and {SooHoo}, Jason and {Tazaki}, Fumie and {Tiede}, Paul and {Tilanus}, Remo P.~J. and {Titus}, Michael and {Toma}, Kenji and {Torne}, Pablo and {Trent}, Tyler and {Trippe}, Sascha and {Tsuda}, Shuichiro and {van Bemmel}, Ilse and {van Langevelde}, Huib Jan and {van Rossum}, Daniel R. and {Wagner}, Jan and {Wardle}, John and {Weintroub}, Jonathan and {Wex}, Norbert and {Wharton}, Robert and {Wielgus}, Maciek and {Wong}, George N. and {Wu}, Qingwen and {Young}, Ken and {Young}, Andr{\'e}},
        title = "{First M87 Event Horizon Telescope Results. I. The Shadow of the Supermassive Black Hole}",
      journal = {\apjl},
         year = 2019,
        month = apr,
       volume = {875},
       number = {1},
          eid = {L1},
        pages = {L1},
          doi = {10.3847/2041-8213/ab0ec7},
archivePrefix = {arXiv},
       eprint = {1906.11238},
 primaryClass = {astro-ph.GA},
       adsurl = {https://ui.adsabs.harvard.edu/abs/2019ApJ...875L...1E}
}

@ARTICLE{2007A&A...475..785M,
       author = {{Meliani}, Z. and {Keppens}, R.},
        title = "{Transverse stability of relativistic two-component jets}",
      journal = {\aap},
         year = 2007,
        month = dec,
       volume = {475},
       number = {3},
        pages = {785-789},
          doi = {10.1051/0004-6361:20078563},
archivePrefix = {arXiv},
       eprint = {0709.3838},
 primaryClass = {astro-ph},
       adsurl = {https://ui.adsabs.harvard.edu/abs/2007A&A...475..785M}
}

@ARTICLE{2009ApJ...705.1594M,
       author = {{Meliani}, Z. and {Keppens}, R.},
        title = "{Decelerating Relativistic Two-Component Jets}",
      journal = {\apj},
         year = 2009,
        month = nov,
       volume = {705},
       number = {2},
        pages = {1594-1606},
          doi = {10.1088/0004-637X/705/2/1594},
archivePrefix = {arXiv},
       eprint = {0910.0332},
 primaryClass = {astro-ph.HE},
       adsurl = {https://ui.adsabs.harvard.edu/abs/2009ApJ...705.1594M}
}

@ARTICLE{2008A&A...491..321M,
       author = {{Meliani}, Z. and {Keppens}, R. and {Giacomazzo}, B.},
        title = "{Faranoff-Riley type I jet deceleration at density discontinuities. Relativistic hydrodynamics with a realistic equation of state}",
      journal = {\aap},
         year = 2008,
        month = nov,
       volume = {491},
       number = {2},
        pages = {321-337},
          doi = {10.1051/0004-6361:20079185},
archivePrefix = {arXiv},
       eprint = {0808.2492},
 primaryClass = {astro-ph},
       adsurl = {https://ui.adsabs.harvard.edu/abs/2008A&A...491..321M}
}

@ARTICLE{2020A&A...635A...5D,
       author = {{Dabhade}, P. and {R{\"o}ttgering}, H.~J.~A. and {Bagchi}, J. and {Shimwell}, T.~W. and {Hardcastle}, M.~J. and {Sankhyayan}, S. and {Morganti}, R. and {Jamrozy}, M. and {Shulevski}, A. and {Duncan}, K.~J.},
        title = "{Giant radio galaxies in the LOFAR Two-metre Sky Survey. I. Radio and environmental properties}",
      journal = {\aap},
         year = 2020,
        month = mar,
       volume = {635},
          eid = {A5},
        pages = {A5},
          doi = {10.1051/0004-6361/201935589},
archivePrefix = {arXiv},
       eprint = {1904.00409},
 primaryClass = {astro-ph.GA},
       adsurl = {https://ui.adsabs.harvard.edu/abs/2020A&A...635A...5D}
}

@ARTICLE{2024Natur.633..537O,
       author = {{Oei}, Martijn S.~S.~L. and {Hardcastle}, Martin J. and {Timmerman}, Roland and {Gast}, Aivin R.~D.~J.~G.~I.~B. and {Botteon}, Andrea and {Rodriguez}, Antonio C. and {Stern}, Daniel and {Calistro Rivera}, Gabriela and {van Weeren}, Reinout J. and {R{\"o}ttgering}, Huub J.~A. and {Intema}, Huib T. and {de Gasperin}, Francesco and {Djorgovski}, S.~G.},
        title = "{Black hole jets on the scale of the cosmic web}",
      journal = {\nat},
         year = 2024,
        month = sep,
       volume = {633},
       number = {8030},
        pages = {537-541},
          doi = {10.1038/s41586-024-07879-y},
archivePrefix = {arXiv},
       eprint = {2411.08630},
 primaryClass = {astro-ph.HE},
       adsurl = {https://ui.adsabs.harvard.edu/abs/2024Natur.633..537O}
}

@ARTICLE{2025A&A...699A.257A,
       author = {{Andernach}, H. and {Br{\"u}ggen}, M.},
        title = "{Properties of giant radio galaxies larger than 3 Mpc}",
      journal = {\aap},
         year = 2025,
        month = jul,
       volume = {699},
          eid = {A257},
        pages = {A257},
          doi = {10.1051/0004-6361/202452961},
archivePrefix = {arXiv},
       eprint = {2505.09181},
 primaryClass = {astro-ph.GA},
       adsurl = {https://ui.adsabs.harvard.edu/abs/2025A&A...699A.257A}
}

@ARTICLE{2019ApJ...874...43L,
       author = {{Lister}, M.~L. and {Homan}, D.~C. and {Hovatta}, T. and {Kellermann}, K.~I. and {Kiehlmann}, S. and {Kovalev}, Y.~Y. and {Max-Moerbeck}, W. and {Pushkarev}, A.~B. and {Readhead}, A.~C.~S. and {Ros}, E. and {Savolainen}, T.},
        title = "{MOJAVE. XVII. Jet Kinematics and Parent Population Properties of Relativistically Beamed Radio-loud Blazars}",
      journal = {\apj},
         year = 2019,
        month = mar,
       volume = {874},
       number = {1},
          eid = {43},
        pages = {43},
          doi = {10.3847/1538-4357/ab08ee},
archivePrefix = {arXiv},
       eprint = {1902.09591},
 primaryClass = {astro-ph.GA},
       adsurl = {https://ui.adsabs.harvard.edu/abs/2019ApJ...874...43L}
}

@ARTICLE{1996A&ARv...7....1C,
       author = {{Carilli}, C.~L. and {Barthel}, P.~D.},
        title = "{Cygnus A}",
      journal = {\aapr},
         year = 1996,
        month = jan,
       volume = {7},
       number = {1},
        pages = {1-54},
          doi = {10.1007/s001590050001},
       adsurl = {https://ui.adsabs.harvard.edu/abs/1996A&ARv...7....1C}
}

@ARTICLE{1993AJ....105.1690F,
       author = {{Fernini}, Ilias and {Burns}, Jack O. and {Bridle}, Alan H. and {Perley}, Rick A.},
        title = "{Very Large Array Imaging of Five Fanaroff-Riley II 3CR Radio Galaxies}",
      journal = {\aj},
         year = 1993,
        month = may,
       volume = {105},
        pages = {1690},
          doi = {10.1086/116547},
       adsurl = {https://ui.adsabs.harvard.edu/abs/1993AJ....105.1690F}
}

@ARTICLE{1997AJ....114.2292F,
       author = {{Fernini}, Ilias and {Burns}, Jack O. and {Perley}, Rick A.},
        title = "{VLA Imaging of Fanaroff-Riley II 3CR Radio Galaxies.II.Eight New Images and Comparisons with 3CR Quasars}",
      journal = {\aj},
         year = 1997,
        month = dec,
       volume = {114},
        pages = {2292},
          doi = {10.1086/118649},
       adsurl = {https://ui.adsabs.harvard.edu/abs/1997AJ....114.2292F}
}

@ARTICLE{2002ApJ...581..948H,
       author = {{Hardcastle}, M.~J. and {Birkinshaw}, M. and {Cameron}, R.~A. and {Harris}, D.~E. and {Looney}, L.~W. and {Worrall}, D.~M.},
        title = "{Magnetic Field Strengths in the Hot Spots and Lobes of Three Powerful Fanaroff-Riley Type II Radio Sources}",
      journal = {\apj},
         year = 2002,
        month = dec,
       volume = {581},
       number = {2},
        pages = {948-973},
          doi = {10.1086/344409},
archivePrefix = {arXiv},
       eprint = {astro-ph/0208204},
 primaryClass = {astro-ph},
       adsurl = {https://ui.adsabs.harvard.edu/abs/2002ApJ...581..948H}
}

@ARTICLE{2002MNRAS.336..328L,
       author = {{Laing}, R.~A. and {Bridle}, A.~H.},
        title = "{Relativistic models and the jet velocity field in the radio galaxy 3C 31}",
      journal = {\mnras},
         year = 2002,
        month = oct,
       volume = {336},
       number = {1},
        pages = {328-352},
          doi = {10.1046/j.1365-8711.2002.05756.x},
archivePrefix = {arXiv},
       eprint = {astro-ph/0206215},
 primaryClass = {astro-ph},
       adsurl = {https://ui.adsabs.harvard.edu/abs/2002MNRAS.336..328L}
}

@ARTICLE{2002MNRAS.336.1161L,
       author = {{Laing}, R.~A. and {Bridle}, A.~H.},
        title = "{Dynamical models for jet deceleration in the radio galaxy 3C 31}",
      journal = {\mnras},
         year = 2002,
        month = nov,
       volume = {336},
       number = {4},
        pages = {1161-1180},
          doi = {10.1046/j.1365-8711.2002.05873.x},
archivePrefix = {arXiv},
       eprint = {astro-ph/0207427},
 primaryClass = {astro-ph},
       adsurl = {https://ui.adsabs.harvard.edu/abs/2002MNRAS.336.1161L}
}

@ARTICLE{2008MNRAS.386..657L,
       author = {{Laing}, R.~A. and {Bridle}, A.~H. and {Parma}, P. and {Feretti}, L. and {Giovannini}, G. and {Murgia}, M. and {Perley}, R.~A.},
        title = "{Multifrequency VLA observations of the FR I radio galaxy 3C 31: morphology, spectrum and magnetic field}",
      journal = {\mnras},
         year = 2008,
        month = may,
       volume = {386},
       number = {2},
        pages = {657-672},
          doi = {10.1111/j.1365-2966.2008.13091.x},
archivePrefix = {arXiv},
       eprint = {0803.2597},
 primaryClass = {astro-ph},
       adsurl = {https://ui.adsabs.harvard.edu/abs/2008MNRAS.386..657L}
}

@ARTICLE{2011MNRAS.417.2789L,
       author = {{Laing}, R.~A. and {Guidetti}, D. and {Bridle}, A.~H. and {Parma}, P. and {Bondi}, M.},
        title = "{Deep imaging of Fanaroff-Riley Class I radio galaxies with lobes}",
      journal = {\mnras},
         year = 2011,
        month = nov,
       volume = {417},
       number = {4},
        pages = {2789-2808},
          doi = {10.1111/j.1365-2966.2011.19436.x},
archivePrefix = {arXiv},
       eprint = {1107.2511},
 primaryClass = {astro-ph.CO},
       adsurl = {https://ui.adsabs.harvard.edu/abs/2011MNRAS.417.2789L}
}

@ARTICLE{2008A&A...488..795R,
       author = {{Rossi}, P. and {Mignone}, A. and {Bodo}, G. and {Massaglia}, S. and {Ferrari}, A.},
        title = "{Formation of dynamical structures in relativistic jets: the FRI case}",
      journal = {\aap},
         year = 2008,
        month = sep,
       volume = {488},
       number = {3},
        pages = {795-806},
          doi = {10.1051/0004-6361:200809687},
archivePrefix = {arXiv},
       eprint = {0806.1648},
 primaryClass = {astro-ph},
       adsurl = {https://ui.adsabs.harvard.edu/abs/2008A&A...488..795R}
}

@ARTICLE{2016A&A...596A..12M,
       author = {{Massaglia}, S. and {Bodo}, G. and {Rossi}, P. and {Capetti}, S. and {Mignone}, A.},
        title = "{Making Faranoff-Riley I radio sources. I. Numerical hydrodynamic 3D simulations of low-power jets}",
      journal = {\aap},
         year = 2016,
        month = nov,
       volume = {596},
          eid = {A12},
        pages = {A12},
          doi = {10.1051/0004-6361/201629375},
archivePrefix = {arXiv},
       eprint = {1609.02497},
 primaryClass = {astro-ph.HE},
       adsurl = {https://ui.adsabs.harvard.edu/abs/2016A&A...596A..12M}
}

@ARTICLE{2019A&A...621A.132M,
       author = {{Massaglia}, S. and {Bodo}, G. and {Rossi}, P. and {Capetti}, S. and {Mignone}, A.},
        title = "{Making Faranoff-Riley I radio sources. II. The effects of jet magnetization}",
      journal = {\aap},
         year = 2019,
        month = jan,
       volume = {621},
          eid = {A132},
        pages = {A132},
          doi = {10.1051/0004-6361/201834512},
archivePrefix = {arXiv},
       eprint = {1812.00657},
 primaryClass = {astro-ph.HE},
       adsurl = {https://ui.adsabs.harvard.edu/abs/2019A&A...621A.132M}
}

@ARTICLE{2022A&A...659A.139M,
       author = {{Massaglia}, S. and {Bodo}, G. and {Rossi}, P. and {Capetti}, A. and {Mignone}, A.},
        title = "{Making Fanaroff-Riley I radio sources. III. The effects of the magnetic field on relativistic jets' propagation and source morphologies}",
      journal = {\aap},
         year = 2022,
        month = mar,
       volume = {659},
          eid = {A139},
        pages = {A139},
          doi = {10.1051/0004-6361/202038724},
archivePrefix = {arXiv},
       eprint = {2112.06827},
 primaryClass = {astro-ph.HE},
       adsurl = {https://ui.adsabs.harvard.edu/abs/2022A&A...659A.139M}
}

@ARTICLE{2024A&A...685A...4R,
       author = {{Rossi}, P. and {Bodo}, G. and {Massaglia}, S. and {Capetti}, A.},
        title = "{The different flavors of extragalactic jets: Magnetized relativistic flows}",
      journal = {\aap},
         year = 2024,
        month = may,
       volume = {685},
          eid = {A4},
        pages = {A4},
          doi = {10.1051/0004-6361/202348864},
archivePrefix = {arXiv},
       eprint = {2402.04707},
 primaryClass = {astro-ph.HE},
       adsurl = {https://ui.adsabs.harvard.edu/abs/2024A&A...685A...4R}
}

@ARTICLE{2021AN....342.1171P,
       author = {{Perucho}, Manel and {L{\'o}pez-Miralles}, Jose and {Reynaldi}, Victoria and {Labiano}, {\'A}lvaro},
        title = "{Jet propagation through inhomogeneous media and shock ionization}",
      journal = {Astronomische Nachrichten},
         year = 2021,
        month = nov,
       volume = {342},
       number = {1171},
        pages = {1171-1175},
          doi = {10.1002/asna.20210051},
archivePrefix = {arXiv},
       eprint = {2109.15234},
 primaryClass = {astro-ph.HE},
       adsurl = {https://ui.adsabs.harvard.edu/abs/2021AN....342.1171P}
}

@ARTICLE{2024A&A...684A..45P,
       author = {{Perucho}, Manel},
        title = "{Shocks, clouds, and atomic outflows in active galactic nuclei hosting relativistic jets}",
      journal = {\aap},
         year = 2024,
        month = apr,
       volume = {684},
          eid = {A45},
        pages = {A45},
          doi = {10.1051/0004-6361/202348624},
archivePrefix = {arXiv},
       eprint = {2401.14218},
 primaryClass = {astro-ph.HE},
       adsurl = {https://ui.adsabs.harvard.edu/abs/2024A&A...684A..45P}
}

@ARTICLE{2007PhRvE..75e6312P,
       author = {{Perucho}, Manuel and {Hanasz}, Michal and {Mart{\'\i}}, Jos{\'e}-Mar{\'\i}a and {Miralles}, Juan-Antonio},
        title = "{Resonant Kelvin-Helmholtz modes in sheared relativistic flows}",
      journal = {\pre},
         year = 2007,
        month = may,
       volume = {75},
       number = {5},
          eid = {056312},
        pages = {056312},
          doi = {10.1103/PhysRevE.75.056312},
archivePrefix = {arXiv},
       eprint = {0705.0441},
 primaryClass = {astro-ph},
       adsurl = {https://ui.adsabs.harvard.edu/abs/2007PhRvE..75e6312P}
}

@ARTICLE{2007A&A...469L..23P,
       author = {{Perucho}, M. and {Lobanov}, A.~P.},
        title = "{Physical properties of the jet in <ASTROBJ>0836+710</ASTROBJ> revealed by its transversal structure}",
      journal = {\aap},
         year = 2007,
        month = jul,
       volume = {469},
       number = {1},
        pages = {L23-L26},
          doi = {10.1051/0004-6361:20077610},
archivePrefix = {arXiv},
       eprint = {0705.0433},
 primaryClass = {astro-ph},
       adsurl = {https://ui.adsabs.harvard.edu/abs/2007A&A...469L..23P}
}

@ARTICLE{2019A&A...627A..79V,
       author = {{Vega-Garc{\'\i}a}, L. and {Perucho}, M. and {Lobanov}, A.~P.},
        title = "{Derivation of the physical parameters of the jet in S5 0836+710 from stability analysis}",
      journal = {\aap},
         year = 2019,
        month = jul,
       volume = {627},
          eid = {A79},
        pages = {A79},
          doi = {10.1051/0004-6361/201935119},
archivePrefix = {arXiv},
       eprint = {1904.02030},
 primaryClass = {astro-ph.HE},
       adsurl = {https://ui.adsabs.harvard.edu/abs/2019A&A...627A..79V}
}

@ARTICLE{2019MNRAS.490.4271M,
       author = {{Matsumoto}, Jin and {Masada}, Youhei},
        title = "{Propagation, cocoon formation, and resultant destabilization of relativistic jets}",
      journal = {\mnras},
         year = 2019,
        month = dec,
       volume = {490},
       number = {3},
        pages = {4271-4280},
          doi = {10.1093/mnras/stz2821},
archivePrefix = {arXiv},
       eprint = {1910.11578},
 primaryClass = {astro-ph.HE},
       adsurl = {https://ui.adsabs.harvard.edu/abs/2019MNRAS.490.4271M}
}

\appendix

\end{document}